%% file: main.tex
\documentclass[onecolumn]{svjour3}
\usepackage{geometry}
\usepackage{graphicx}
\usepackage{latexsym}
\usepackage{amsmath}
\usepackage{amssymb}
\usepackage{amsfonts}
\usepackage{braket}
\usepackage{physics}
\usepackage{slashed}
\usepackage{float}
\usepackage{hyperref}
\usepackage{circuitikz}
\usepackage{longtable}
\usepackage{caption}
\usepackage[
    backend=biber,
    bibstyle=numeric,               
    hyperref=true,
    sorting=none,
    style=phys
]{biblatex}

\AtEveryBibitem{\clearlist{language}}

\begin{document}

\title{Delayed Choice Lorentz Transformations on a Qubit}

\author{Lucas Burns, Sacha Greenfield, Justin Dressel}

\institute{L. Burns, S. Greenfield, J. Dressel \at
    Institute for Quantum Studies, \\ 
    Schmid College of Science and Technology, \\
    Chapman University, Orange CA 92866, USA. 
    \and
    S. Greenfield \at 
    Center for Quantum Information Science and Technology, \\
    Department of Physics and Astronomy, \\
    University of Southern California, Los Angeles, CA 90089, USA.}
\date{}

\maketitle

\begin{abstract}
A continuously monitored quantum bit (qubit) exhibits competition between unitary Hamiltonian dynamics and non-unitary measurement-collapse dynamics, which for diffusive measurements form an enlarged transformation group equivalent to the Lorentz group of spacetime. We leverage this equivalence to develop a four-dimensional generalization of the three-dimensional Bloch ball to visualize the state of a monitored qubit as the four-momentum of an effective classical charge affected by a stochastic electromagnetic force field. Unitary qubit dynamics generated by Hermitian Hamiltonians correspond to elliptic spatial rotations of this effective charge while non-unitary qubit dynamics generated by non-Hermitian Hamiltonians or stochastic measurement collapse correspond to hyperbolic Lorentz boosts. Notably, to faithfully emulate the stochastic qubit dynamics arising from continuous qubit measurement, the stochastic electromagnetic fields must depend on the velocity of the charge they are acting on. Moreover, continuous qubit measurements admit a dynamical delayed choice effect where a future experimental choice can appear to retroactively determine the type of past measurement backaction, so the corresponding point charge dynamics can also exhibit delayed choice Lorentz transformations in which a future experimental choice determines whether stochastic force fields are electric or magnetic in character long after they interact with the particle.
\end{abstract}

\tableofcontents

\section{Introduction}

Recent experimental progress in quantum hardware has enabled concurrent monitoring of quantum state evolution, highlighting the inherent tension between unitary Hamiltonian dynamics and non-unitary measurement-collapse dynamics. Unlike typical prepare-and-measure scenarios that relegate the use of measurement solely to the beginning and end of an experimental protocol, monitoring the state evolution irreducibly interleaves Hamiltonian dynamics with measurement-collapse dynamics to produce \emph{conditioned quantum state trajectories} as the most complete description of the system given the monitored record \cite{Gardiner1985,barchielli1986measurement,diosi1988continuous,belavkin1992quantum,dalibardWavefunctionApproachDissipative1992,carmichaelOpenSystemsApproach1993,WisemanMilburnTrajectories1993,goetschLinearStochasticWave1994,wisemanQuantumTrajectoriesQuantum1996,korotkov2001selective,JacobsIntroContMeas06,Gambetta2006,Gambetta2008,Korotkov2011-qbayes,Korotkov2016,WisemanBook}. Superconducting quantum processors in particular have demonstrated a range of utility for such quantum trajectories, including to monitor quantum jumps \cite{Vijay2011-vn,Hatridge2013,Sun2014-zj,Vool2016-de}, passively track diffusive dynamics \cite{Murch2013,Weber2014,Campagne-IbarcqPRX2016,Ficheux2018,Hacohen-Gourgy2016,garcia-pintosObservableDynamicsContinuously2017,FlurinPRX2020,koolstra2022monitoring}, actively control dynamics with feedback \cite{Vijay2012-rabistable,de2014reversing,Dressel2017,Minev2019-gu,Martin2020,Hacohen-Gourgy_and_Martin,hacohen2018incoherent,kumari2023qubit,lewalle2024optimal,greenfield2024stabilizing,greenfield2025unified}, remotely generate entanglement \cite{Roch2014,Chantasri2016-entangledstats}, and continuously check syndromes for error correction \cite{Mohseninia2020,Atalaya2020,ChenPRR2020,livingston2022experimental}.

The effective dynamics resulting from such concurrent monitoring are fundamentally non-unitary and stochastic, with a richer mathematical structure than standard unitary quantum dynamics. In the ideal case, efficiently monitoring a quantum system (i.e., with no dissipative information loss) produces evolution generated by stochastic non-Hermitian Hamiltonians that can have interesting properties not shared by Hermitian Hamiltonians \cite{gisinSimpleNonlinearDissipative1981, plenioQuantumjumpApproachDissipative1998, brodyMixedStateEvolutionPresence2012, sergiLinearQuantumEntropy2016,alipour2020shortcuts,garcia2022unifying,chen2022decoherence,matsoukas2023non,karmakar2025noise}. Indeed, the recent interest in engineering quasi-deterministic non-Hermitian Hamiltonians \cite{harrington2022engineered} (e.g., via post-selection on a sequence of no-jump results given a monitoring process with rare jumps \cite{naghilooQuantumStateTomography2019, ashida2020, chenQuantumJumpsNonHermitian2021, roccatiNonHermitianPhysicsMaster2022, wakefieldNonHermiticityQuantumNonlinear2024, martinez-azconaQuantumDynamicsStochastic2025}) is a special case of such monitored evolution that underscores the potential benefits of studying the enlarged class of transformations inherent to monitored dynamics. 

For the case of diffusive monitoring, which is the standard readout protocol for superconducting circuits \cite{Gambetta2008,Korotkov2011-qbayes,Korotkov2016}, each observation in a temporal sequence yields a non-projective (weak) measurement transformation \cite{aharonov_how_1988,Dressel-2014e} on the state that is technically invertible with a complementary observation, implying that the enlarged class of monitored dynamics still forms a (non-unitary) transformation group. The lower likelihood of observing inverse transformations produces the apparent statistical arrow of time that favors collapse toward measurement eigenstates on average while retaining invertibility \cite{dresselArrowTimeContinuous2017}. Focusing on qubits as the simplest nontrivial quantum system, the class of transformations for such diffusively monitored dynamics widens from the special unitary group $\text{SU}(2)$ to the special linear group $\text{SL}(2,\mathbb{C})$ \cite{jordanQubitFeedbackControl2006}. However, this enlarged group representation is contingent on using the state norm to partially track the accumulated probability for the observed measurement record, such that the number of independent real parameters needed to characterize the qubit state increases from 3 to 4. This extra parameter is problematic for the traditional Bloch ball visualization of a qubit, which assumes a normalized state that can be identified with a 3-dimensional spin-vector. Thus, traditional analyses of monitored qubit dynamics have renormalized the state after each observation to yield nonlinear dynamics that restore the viability of the Bloch ball visualization at the cost of obfuscating the enlarged group structure.

In this paper, we generalize the 3-dimensional Bloch ball to a 4-dimensional representation that accommodates the enlarged group structure of monitored qubit dynamics. We find that this 4-dimensional representation naturally has the Lorentz group structure of spacetime, which may be somewhat surprising since an abstract qubit has no connection to spacetime \emph{a priori}. Nevertheless, the enlarged general linear group $\text{SL}(2,\mathbb{C})$ of monitored qubit dynamics is equivalent to a matrix representation of the Lorentz group of spacetime transformations suitable for Weyl spinors \cite{dreiner2010two} and twistors \cite{penroseSpinorsSpaceTimeVolume1984}, and naturally relates to Clifford-algebraic treatments of spacetime symmetries \cite{hestenes2003spacetime, doranStatesOperatorsSpacetime1993a,doranLieGroupsSpin1993,doranGeometricAlgebraPhysicists2007,baylisComplexAlgebraPhysical2012,dressel_spacetime_2015,burnsAcousticElectromagneticField2020a,burns2024spacetime}. As such, our representation yields a precise correspondence between the four real components of the unnormalized qubit state and the components of a proper four-momentum for a point charge in spacetime (and generalizes Ref.~\cite{botetDualityNonHermitianTwostate2019}). The three-dimensional Bloch vector obtained after renormalization then corresponds to the three-velocity of that point charge in a particular frame, providing an intriguing perspective on the significance of state renormalization during monitoring. With this spacetime representation, the enlarged group of stochastic dynamics for the monitored qubit state intuitively corresponds to the dynamics of a spacetime point charge being influenced by stochastic electromagnetic force fields \cite{boyerGeneralConnectionRandom1975, goedeckeStochasticElectrodynamicsStochastic1983, huangDiscreteExcitationSpectrum2015, drummondQfunctionsModelsPhysical2020}, which gives a concrete and familiar way to visualize monitored qubit dynamics beyond the Bloch ball. Magnetic fields correspond to Hermitian generators of elliptic (unitary) dynamics producing spatial rotations of the charge, while electric fields correspond to anti-Hermitian generators of hyperbolic (measurement-collapse) dynamics producing linear boosts of the charge. 

Much like one can interpret the three-dimensional Bloch vector representation of a normalized qubit state as a classical spin hidden variable model, one can also interpret our generalization to a spacetime point charge as a concrete classical hidden variable model for a monitored quantum state. One thus expects that any nonclassical features of a qubit should manifest as strange features of its corresponding spacetime charge. We show that the stochastic dynamics of a monitored qubit indeed have several strange features that one would not expect for a truly classical charge in spacetime. First, the corresponding stochastic electromagnetic fields that influence the charge motion must depend on the velocity of the charge they are acting on to reproduce the behavior of measurement collapse. Such a velocity-dependent external field is a form of fine-tuned and instantaneous feedback control that would be highly unusual and difficult to experimentally arrange for an actual point charge. Second, more detailed analysis of how the qubit measurement occurs reveals an intrinsically retrocausal character hidden in the controllable parameters of the corresponding force fields \cite{carmichaelOpenSystemsApproach1993}: both the type of electromagnetic field and their specific fluctuations can be determined at later times than their interaction with the charge. In particular, an experimenter is free to vary an angle $\theta(t)$ that determines the type of electromagnetic field, and thus the type of Lorentz transformation, which affects the charge at an earlier interaction time $t-\delta t_q$. Such a dynamical delayed choice effect results from the entanglement of the qubit with the detecting apparatus and thus cannot be replicated by causal electromagnetic fields affecting a truly classical charge. Our four-dimensional representation for the state of a monitored qubit thus not only serves as a useful visualization that preserves the natural group structure of its dynamics, but also helps clarify which features of those dynamics have nonclassical character. In this way our qubit representation is similar in spirit to other toy models that have been used to identify specific nonclassical features through contrast, such as Bell-nonlocality \cite{bellEinsteinPodolskyRosen1964b,wisemanSteeringEntanglementNonlocality2007, wittmannLoopholefreeEinsteinPodolsky2012}, Leggett-Garg invasive measurability \cite{KorotkovLG,Jordan2006,Dressel2011,Dressel2014,White2016}, and contextuality     \cite{spekkensContextualityPreparationsTransformations2005,Spekkens2007, chiribellaQuantumTheoryInformational2016,garcia-pintosObservableDynamicsContinuously2017}.

The paper is organized as follows: In Section~\ref{sec:review}, we briefly review how to model quantum state trajectories for a qubit. In Section~\ref{sec:kinematics}, we develop the kinematical correspondence between a point charge in spacetime and an unnormalized qubit state. In Section~\ref{sec:dynamics}, we examine the deterministic dynamics of a classical point charge and show the equivalence between the electromagnetic Lorentz force and the non-Hermitian dynamics of an unnormalized qubit state. In Section~\ref{sec:stochastics}, we examine the stochastic dynamics of a point charge in a fluctuating electromagnetic field and highlight the unusual constraints on the fields that are required to achieve correspondence with the stochastic dynamics of a monitored qubit. In Section~\ref{sec:delay}, we give a more detailed derivation of the measurement process for a superconducting qubit to highlight the delayed choice nature of the effective measurement parameters and the implications for the corresponding stochastic electromagnetic fields affecting its analogous point charge. We conclude in Section~\ref{sec:conclusion}. 

\section{Qubits, Measurement, and Trajectories}\label{sec:review}

To establish the foundation for a formal correspondence between point charge and qubit, we first review the mathematical framework of state transformations and measurement trajectories. Here we focus on two-level (qubit) systems, for which normalized pure states have a complex vector representation,
\begin{align}
    \ket \psi &= c_0 \ket 0 + c_1 \ket 1, & c_0,c_1&\in\mathbb{C}, & \braket{\psi}{\psi} = 1,
\end{align}
and for which normalized convex mixtures $\hat{\rho} = \sum_m p_m \ket{\psi_m}\!\bra{\psi_m}$ with $p_m\in[0,1]$ such that $\sum_m p_m = 1$, have a matrix representation,
\begin{align}
    \hat{\rho} &= \frac{1}{2}\left(\hat{1} + \sum_{k=1}^3 S_k\,\hat{\sigma}_k\right), & \text{Tr}(\hat{\rho}) &= 1, & S_k &= \text{Tr}(\hat{\sigma}_k\hat{\rho})\in\mathbb{R},
\end{align}
in terms of the Pauli matrices defined as
\begin{subequations}
\begin{align}
    \hat 1 &= \hat \sigma_0 = \ketbra{0}{0} + \ketbra{1}{1}, &
    \hat \sigma_x &= \hat \sigma_1 = \ketbra{0}{1} + \ketbra{1}{0}, \\
    \hat \sigma_y &= \hat \sigma_2 = -i (\ketbra{0}{1} - \ketbra{1}{0}), &
    \hat \sigma_z &= \hat \sigma_3 = \ketbra{0}{0} - \ketbra{1}{1}.
\end{align}
\end{subequations}
The Pauli operator expectation values $S_k$ fully parametrize a mixed state as components of an effective spin (Bloch) vector $\vec{S} = (S_1,S_2,S_3)\in\mathbb{R}^3$ satisfying $|\vec{S}|^2 \leq 1$, making a general normalized qubit state space equivalent to a filled three-dimensional unit (Bloch) ball.

Quantum states in the Schrödinger picture undergo two fundamental types of transformations: (i) unitary Hamiltonian evolution, and (ii) non-unitary collapse from measurement backaction. Unitary transformations satisfy $\hat U \hat U^\dagger = \hat U^\dagger \hat U = \hat 1$ and take the form
\begin{align}\label{eq:unitary}
    \hat U &= \exp\!\left[-i \left(\phi_0 \hat 1 + \sum_{k=1}^3 \phi_k \hat \sigma_k\right)/2\right], & \ket \psi &\mapsto \hat{U}\ket\psi, & \hat{\rho} \mapsto \hat{U}\hat{\rho}\hat{U}^\dagger,
\end{align}
where $\phi_\mu \in \mathbb R$ for $\mu = 0,1,2,3$. Since $\phi_0$ changes only the global phase, it is a gauge freedom for an isolated qubit, so is usually omitted. For an efficient measurement (i.e., without information loss) of a particular result $r$ of a detector, the quantum state transforms according to the non-unitary collapse backaction
\begin{align}
    \ket \psi &\mapsto \frac{\hat M_r \ket \psi}{\sqrt{p(r)}}, & \hat{\rho} &\mapsto \frac{\hat{M}_r\hat{\rho}\hat{M}_r^\dagger}{p(r)}, 
\end{align}
where
\begin{align}\label{eq:povm}
    p(r) &= \bra \psi \hat M_r^\dagger \hat M_r \ket \psi, & p(r) &= \Tr(\hat{M}_r^\dagger\hat{M}_r\hat{\rho}),
\end{align}
is the probability of the result $r\in R$. To ensure probability normalization $\int_R dr\, p(r) = 1$, the measurement (Kraus) operators $\hat M_r$ satisfy the completeness relation $\int_R dr \hat M_r^\dagger \hat M_r = \hat 1$, making $dr\,\hat P_r \equiv dr\,\hat M_r^\dagger \hat M_r$ a positive operator-valued measure (POVM) over the detector result index $r\in R$.

This measurement collapse can be understood as a quantum generalization of Bayes' rule \cite{Korotkov2011-qbayes,Korotkov2016,Dressel-2014e}. For specificity, consider a measurement in the $\hat \sigma_z$ basis. The simplest Bayesian update of the state $\ket \psi = \sqrt{p(0)} \ket 0 + \sqrt{p(1)} e^{i \varphi} \ket 1$ given a measured result $r$ is described by a measurement operator of the form
\begin{align}\label{eq:M_r0}
    \hat M_r = \sqrt{p(r|0)} \ketbra{0}{0} + \sqrt{p(r|1)} \ketbra{1}{1} 
\end{align}
where $p(r|0)$ and $p(r|1)$ are the empirical detector response likelihoods for obtaining the result $r$ given definite preparations of $\ket 0$ and $\ket 1$ states, respectively, ensuring that updated probabilities in the normalized post-measurement state adhere to Bayes' rule,
\begin{align}\label{eq:post-measurement-state}
    \ket{\psi_r} &= \frac{\hat M_r \ket \psi}{\norm{\hat M_r \ket \psi}} = \sqrt{p(0|r)} \ket 0 + \sqrt{p(1|r)} e^{i \varphi} \ket 1, & p(k|r) &= \frac{p(r|k)p(k)}{p(r|0)p(0) + p(r|1)p(1)}.
\end{align}
Notably, the measurement operator in Eq.~\eqref{eq:M_r0} can be written in the form,
\begin{align}\label{eq:M_r}
    \hat{M}_r &= \sqrt{\bar{p}(r)} \exp\left( \hat \sigma_z \lambda(r)/4\right), & \bar{p}(r) &\equiv \sqrt{p(r|0) p(r|1)}, & \lambda(r) &\equiv \ln \frac{p(r|0)}{p(r|1)},
\end{align}
involving the geometric mean $\bar{p}(r)$ of the detector likelihoods and the surprisal difference $\lambda(r)$ \cite{jaynesProbabilityTheoryLogic2003}. The mean likelihood $\bar{p}(r)$ is largest at maximal overlap and smallest at minimal overlap, bounded above by the largest and below by the smallest of the two probabilities, and it can be understood as a measure of indistinguishability. In contrast, the surprisal difference $\lambda(r)$ is the logarithmic Bayes factor, a signed measure of distinguishability that fully determines the measurement backaction on the state. The limits of $\lambda(r)\to\pm\infty$ are projective measurements, while $\lambda(r)\approx 0$ are weak measurements \cite{aharonov_how_1988}, and other values of $\lambda(r)$ are generalized measurements of intermediate strength \cite{kraus_states_1983}. The mean likelihood $\bar{p}(r)$ contributes a trivial global rescaling of the state and cancels in the renormalization.

To simplify the description of the measurement operator, we use a different normalization convention that keeps only the essential information $\lambda(r)$ in the operator and tracks the mean likelihood $\bar{p}(r)$ as part of the probability definition separately,
\begin{align}\label{eq:L_r}
    \hat{L}_r &\equiv \frac{\hat{M}_r}{\sqrt{\bar{p}(r)}},
    & \ket{\widetilde{\psi}_r} &\equiv \hat{L}_r\ket{\psi}, & \hat{s}_r &\equiv \hat{L}_r\hat{\rho}\hat{L}_r^\dagger.
\end{align}
This choice of scaling retains minimal information in a linear update to an unnormalized conditioned state, $\ket{\widetilde{\psi}_r}$ or $\hat{s}_r$, while still allowing recovery of the measurement probability,
\begin{align}\label{eq:p_r}
      p(r) &= \bar{p}(r)\,\braket{\widetilde{\psi}_r}, &  p(r) &= \bar{p}(r)\,\Tr(\hat{s}_r),
\end{align}
using both the state norm and the separately tracked mean likelihood $\bar{p}(r)$. The trace of the unnormalized state after a single measurement update is
\begin{align}\label{eq:strace}
    \Tr(\hat s_r) &= \frac{p(r)}{\bar{p}(r)} = e^{\lambda(r)/2} p(0) + e^{-\lambda(r)/2} p(1),
\end{align}
which tells us how informative the single outcome $r$ was given our prior knowledge $p(0), p(1)$ of the state, irrespective of the information conveyed by $r$. Whereas the surprisal $\lambda(r)$ (Eq.~\ref{eq:M_r}) specifies the distinguishability of the measurement along with the direction of information gain, $\operatorname{Tr}(\hat s_r)$ specifies only how much the measurement distinguished the state in the measurement basis without specifying what was distinguished. In the extreme limit of maximally informative measurement where $p(r|0)$ and $p(r|1)$ limit to delta functions at $\pm 1$, the surprisal $\lambda(r) \mapsto \pm \infty$ while $\operatorname{Tr}(\hat s_r)$ limits to $+\infty$ in both cases, indicating collapse to one of the definite states without specifying which.

For a sequence of measurements, this linear update dramatically simplifies bookkeeping \cite{goetschLinearStochasticWave1994, wisemanQuantumTrajectoriesQuantum1996,Dressel2017}; for example,
\begin{align}
    \hat{\rho} &\xrightarrow{(r_1,r_2)} \hat{s}_{r_1,r_2} \equiv (\hat{L}_{r_2}\hat{L}_{r_1})\hat{\rho}(\hat{L}_{r_2}\hat{L}_{r_1})^\dagger, & p(r_1,r_2) &= \bar{p}(r_1)\bar{p}(r_2)\Tr(\hat{s}_{r_1,r_2}).
\end{align}
For measurements in other bases, the measurement operators in Eqs.~\eqref{eq:M_r} and \eqref{eq:L_r} generalize to 
\begin{align}\label{eq:measurement}
    \hat M_r &= \exp\!\left[\left(\alpha_0(r) \hat 1 + \sum_{k=1}^3\alpha_k(r) \hat \sigma_k\right)/2\right], &
    \hat L_r &= \exp\!\left[\sum_{k=1}^3\alpha_k(r) \hat \sigma_k/2\right],
\end{align}
where $\alpha_\mu(r) \in \mathbb R$ for $\mu=0,1,2,3$, which directly parallels Eq.~\eqref{eq:unitary} for unitary evolution. Analogously to the global phase $\phi_0$ in the unitary case, $\alpha_0(r) = \ln\bar{p}(r)$ contributes only to the scaling of the state before normalization, so is omitted for the minimal operator $\hat{L}_r$.

The two types of transformations, unitary in Eq.~\eqref{eq:unitary} and measurement in Eqs.~\eqref{eq:L_r} and \eqref{eq:measurement}, have fundamentally different geometric properties that will prove central to the correspondence developed in the next section. For example, unitary evolution generated by $\hat \sigma_z$ is elliptic,
\begin{align}
    e^{-i \hat \sigma_z \phi/2} = \hat 1 \cos (\phi/2) - i \hat \sigma_z \sin (\phi/2),
\end{align}
while backaction generated by measuring $\hat \sigma_z$ is hyperbolic,
\begin{align}
    e^{\hat \sigma_z \alpha/2} = \hat 1 \cosh (\alpha/2) + \hat \sigma_z \sinh (\alpha/2).
\end{align}
Notably, hyperbolic transformations yield asymptotic fixed points, unlike elliptic transformations. That is, if one were to obtain the same measurement outcome $r$ with the same strength $\alpha(r)$ repeatedly, then the state would converge to a fixed measurement eigenstate. For example, $\alpha(r)>0$ would yield,
\begin{align}
    \lim_{n \to \infty} \frac{\hat L_r^n \ket \psi}{\norm{\hat L_r^n\ket \psi}} = \lim_{n \to \infty} \frac{e^{n \hat \sigma_z \alpha(r)/2}}{\sqrt{p(0) e^{n\alpha(r)} + p(1) e^{-n\alpha(r)}}} \ket \psi 
    = \lim_{n \to \infty} (\ketbra 0 + e^{-n \alpha(r)} \ketbra 1) \frac{\ket \psi}{\sqrt{p(0)}} 
    = \ket 0,
\end{align}
while $\alpha(r) < 0$ would yield the fixed point $\ket{1}$. 
In stark contrast, repeated unitary rotations are periodic and never converge to a fixed point.

In addition to the informational backaction $\hat{L}_r$, measurement can also include non-informational backaction in the form of conditioned unitary dynamics $\hat{U}_r$, yielding a composite transformation,
\begin{align}\label{eq:R_r}
    \hat{R}_r &\equiv \hat{U}_r\hat{L}_r,
    & \ket{\widetilde{\psi}_r} &\equiv \hat{R}_r\ket{\psi}, & \hat{s}_r &\equiv \hat{R}_r\hat{\rho}\hat{R}_r^\dagger,
\end{align}
with the same probability as in Eq.~\eqref{eq:p_r}.
This general form for an efficient measurement operator thus has an exponential representation weighting the Pauli operators with complex parameters,
\begin{align}\label{eq:slR}
    \hat R_r &= \exp\left(\sum_{k=1}^3 \zeta_k(r) \hat \sigma_k/2 \right), & \zeta_k(r) &\equiv \alpha_k(r) - i\phi_k(r) \in\mathbb{C},
\end{align}
and is thus an element of the closed special linear group $\text{SL}(2, \mathbb C)$. 

Sequences of measurements yield quantum state trajectories \cite{Gardiner1985,barchielli1986measurement,diosi1988continuous,belavkin1992quantum,dalibardWavefunctionApproachDissipative1992,carmichaelOpenSystemsApproach1993,WisemanMilburnTrajectories1993,goetschLinearStochasticWave1994,wisemanQuantumTrajectoriesQuantum1996,korotkov2001selective,JacobsIntroContMeas06,Gambetta2006,Gambetta2008,Korotkov2011-qbayes,Korotkov2016,WisemanBook} that depend on the entire history $(r_1, \ldots, r_n)$ of measurement results, 
\begin{align}
    \ket{\widetilde \psi_{r_1,r_2,\ldots,r_n}} &= \hat R_{r_n} \cdots \hat R_{r_2} \hat R_{r_1} \ket{\psi}, &
    \hat{s}_{r_1,r_2,\ldots,r_n} &= (\hat R_{r_n} \cdots \hat R_{r_2} \hat R_{r_1})\hat{\rho}(\hat{R}_{r_n}\cdots \hat R_{r_2} \hat R_{r_1})^\dagger,
\end{align}
with corresponding trajectory probabilities,
\begin{subequations}
\begin{align}
    p(r_1,r_2,\ldots,r_n) &= [\bar{p}(r_1)\bar{p}(r_2)\cdots\bar{p}(r_n)]\braket{\widetilde \psi_{r_1,r_2,\ldots,r_n}}, \\
    p(r_1,r_2,\ldots,r_n) &= [\bar{p}(r_1)\bar{p}(r_2)\cdots\bar{p}(r_n)]\Tr(\hat{s}_{r_1,r_2,\ldots,r_n}).
\end{align}
\end{subequations}

A continuous-time formulation of such a quantum state trajectory can be formally obtained by interpreting the complex $\zeta_k(r)$ weighting the generators $\hat{\sigma}_k$ as integrations over the time interval $\Delta t$ between each measurement, $\zeta_k(r) = \int_0^{\Delta t}[\gamma_k(t) - i \omega_k(t)]dt$, with the observed stochastic results $\zeta_k(r)$ interpreted as averages over $\Delta t$ of finer-grained stochastic rates $\gamma_k(t)$ and $\omega_k(t)$. When this formal interpolation is possible, the state trajectory is effectively generated by a stochastic non-Hermitian time-dependent Hamiltonian $\hat{R}_r = \exp(-i\int_0^{\Delta t}\hat{H}_{\rm eff}(t)dt/\hbar)$ with $\hat{H}_{\rm eff}(t) = \sum_k \hbar(\omega_k(t) + i\gamma_k(t))\hat{\sigma}_k/2$. Notably, deterministic non-Hermitian Hamiltonians are a special case of this sort of continuous-time interpolation.

We focus on a Gaussian measurement as a key example of a stochastic qubit state trajectory that admits a continuous-time interpolation \cite{JacobsAndSteck2006}, where the result $r$ for each interval $\Delta t$ is sampled from conditional Gaussian distributions centered at distinct means $\pm\bar{r}$ for each measured qubit state,
\begin{align}
    p(r|0) &= \sqrt{\frac{\Gamma \Delta t}{\pi}}\exp(-\Gamma \Delta t(r - \bar{r})^2), &
    p(r|1) &= \sqrt{\frac{\Gamma \Delta t}{\pi}}\exp(-\Gamma \Delta t(r + \bar{r})^2).
\end{align}
Since the variance of each distribution, $1/(2\Gamma\Delta t)$, inversely depends on the time interval $\Delta t$, the average $\bar r = \sum_{k=1}^n r_k/n$ of a sequence of stochastic results $(r_1,\ldots,r_n)$ will also be Gaussian-distributed with a variance $1/(2\Gamma\,n\Delta t)$ that inversely depends on the total time interval $n \Delta t$. This variance scaling permits an interpolation to a continuum limit, where each discrete $r$ observed over a finite $\Delta t$ corresponds to a formal integral over a white-noise process characterized by the single rate parameter $\Gamma$. This scaling also further clarifies interpretation of $\operatorname{Tr}(\hat s_r)$ in Eq.~\ref{eq:strace}: a sequence of $n$ weaker measurements with strength $\Gamma$ integrates to a single, stronger measurement with strength $n \Gamma$, which is consistent with calculation of the unnormalized state norm due to a sequence of $n$ measurements:
\begin{equation}
    \operatorname{Tr}(\hat s_{(r_1,\ldots,r_n)}) = e^{\sum_{k=1}^n \lambda(r_k)} \, p(0) + e^{-\sum_{k=1}^n \lambda(r_k)}  \, p(1) = e^{\bar{\lambda}(\bar r)} p(0) + e^{-\bar{\lambda}(\bar r)} p(1), 
\end{equation}
where $\bar \lambda$ is the modified surprisal given the conditional Gaussian distributions modified with $\Gamma \Delta t \rightarrow \Gamma (n \Delta t)$ and $\bar r$ is the sequence average defined previously. Thus, we may understand the limit of infinitely strong measurement where $\lambda(r) \rightarrow \pm \infty$ as the integration of an infinite sequence of finite strength measurements, in which $\lim_{n \rightarrow \infty}\operatorname{Tr}(\hat s_{(r_1,\ldots,r_n)}) = +\infty$ coincides with perfect collapse to a measurement eigenstate.  

When $\pm\bar{r} = \pm1$, the observed result $r$ is scaled to the eigenvalues of the monitored observable $\hat{\sigma}_z$, in which case the parameter $\Gamma$ represents the \emph{measurement-dephasing rate} observed for the ensemble-averaged dynamics,
\begin{align}\label{eq:ensemble-dephasing}
    \hat{\rho}(t + \Delta t) &= \int \hat{M}_r\hat{\rho}\hat{M}_r^\dagger\,dr = \frac{1}{2}\left(\hat{1} + e^{-\Gamma \Delta t}\left[S_x(t)\hat{\sigma}_x + S_y(t)\hat{\sigma}_y\right] + S_z(t)\hat{\sigma}_z\right).
\end{align}

Importantly, as will be discussed further in Section~\ref{sec:delay}, a fixed measurement-dephasing rate $\Gamma$ is also compatible with a wider class of Gaussian measurements that include both non-unitary collapse backaction and unitary phase-jitter backaction. The ensemble-average dephasing rates from each type of backaction sum in a complementary way to yield the total dephasing rate $\Gamma$. Choosing an angle-dependent scaling $\bar{r}(\theta) = \cos\theta$ for the qubit-state-dependent means $\pm\bar{r}(\theta)$ yields such a Gaussian measurement operator with complementary contributions jointly satisfying Eq.~\eqref{eq:ensemble-dephasing},
\begin{subequations}\label{eq:gaussian-theta}
\begin{align}
    \hat{M}_{r,\theta} &= \sqrt{\bar{p}(r,\theta)}\,\hat{U}_{r,\theta}\hat{L}_{r,\theta}, & 
    \bar{p}(r,\theta) &= \sqrt{\frac{\Gamma \Delta t}{\pi}}\exp(-\Gamma\Delta t(r^2 + \cos^2\theta)), \\
    \hat{U}_{r,\theta} &= \exp(-i \Gamma \Delta t\,r\sin\theta\,\hat{\sigma}_z), & 
    \hat{L}_{r,\theta} &= \exp(\Gamma \Delta t\,r\cos\theta\,\hat{\sigma}_z).
\end{align}
\end{subequations}
The contribution $\hat{L}_{r,\theta}$ represents the informational measurement collapse backaction, while the contribution $\hat{U}_{r,\theta}$ represents the complementary non-informational unitary phase-jitter backaction. Section~\ref{sec:delay} will clarify that the parameter $\theta$ physically corresponds to the choice of detector basis that is measured long after the detector interacts with the qubit: the interaction fixes the ensemble dephasing rate $\Gamma$ for the qubit but does not pre-determine the post-interaction detector-basis choice corresponding to $\theta$. After rescaling by the $\sqrt{\Delta t}$-dependent geometric mean probability $\bar{p}(r,\theta)$, the composite transformation has the exponential form of Eq.~\eqref{eq:slR},
\begin{align}\label{eq:theta}
    \hat{R}_{r,\theta} &= \frac{\hat{M}_{r,\theta}}{\sqrt{\bar{p}(r,\theta)}} = \hat{U}_{r,\theta}\hat{L}_{r,\theta} = \exp(\Gamma \Delta t\,r e^{-i\theta}\,\hat{\sigma}_z) = \exp(\zeta(r,\theta)\hat{\sigma}_z/2) \in \text{SL}(2,\mathbb{C}),
\end{align}
in terms of a complex angle $\zeta(r,\theta) = 2\Gamma\Delta t\,r e^{-i\theta}$ that is linear in $\Delta t\,r$ \cite{Dressel-2014e}. The continuum limit to a white noise process can then be taken, formally yielding for each infinitesimal time step $dt$ an observed signal $\Delta t\,r \to dt\,r = dt\,\cos\theta\,\text{Tr}(\hat{\sigma}_z\,\hat{\rho}(t)) + dW/\sqrt{2\Gamma}$ that approximately tracks the mean of $\cos\theta\,\hat{\sigma}_z$ in the state $\hat{\rho}(t)$ with additive zero-mean white noise described by a stochastic Wiener increment $dW$ with variance $dt$. Expanding the normalized state update $\hat{\rho}(t + dt) = \hat{R}_{r,\theta}\hat{\rho}(t)\hat{R}^\dagger_{r,\theta}/\text{Tr}[\hat{R}^\dagger_{r,\theta}\hat{R}_{r,\theta}\hat{\rho}(t)]$ to linear order in $dt$ using the It\^o rule $dW^2 = dt$ then yields the It\^o picture (forward-difference) stochastic differential equation,
\begin{align}\label{eq:stochastic-me}
    d\hat{\rho}(t) &= \frac{\Gamma}{2}\left(\hat{\sigma}_z\hat{\rho}(t)\hat{\sigma}_z - \hat{\rho}(t)\right)dt + \sqrt{2\Gamma}\left(\frac{e^{-i\theta}\hat{\sigma}_z\hat{\rho}(t) + e^{i\theta}\hat{\rho}(t)\hat{\sigma}_z}{2} - \cos\theta\Tr(\hat{\sigma}_z\,\hat{\rho}(t)) \hat \rho(t) \right)dW,
\end{align}
with the $\theta$-independent ensemble-averaged dynamics in the expected form of a Lindblad master equation ($dW\to 0$), modified by a $\theta$-dependent innovation term linear in the zero-mean additive white noise $dW$. 

\section{Kinematical Correspondence}\label{sec:kinematics}

The group $\text{SL}(2, \mathbb C)$ appearing in Eq.~\eqref{eq:slR} is the spinor representation of the (proper orthochronous) Lorentz group $\text{SO}^+(1,3)$ \cite{doranLieGroupsSpin1993} and establishes a direct correspondence between spacetime structure and qubit transformations. That is, unitary Hamiltonian evolution has the same mathematical structure as a spatial rotation, while the measurement of a detector result $r$ has the same mathematical structure as a Lorentz boost (after factoring out the mean probability $\bar{p}(r)$ from the transformation of the unnormalized state $\hat{s}_r$).

To emphasize this point, consider the simple example of a spacetime particle with rest-mass $m$ and four-momentum $\underline{p}$. Starting the particle initially at rest, a Lorentz boost yields,
\begin{align}
    \underline{p} = (mc,\, \vec 0) \qquad &\longrightarrow \qquad \underline{p}' = (\gamma mc,\, \gamma mc \vec \beta),
\end{align}
where $\vec \beta = \vec v/c = (v_1, v_2, v_3)/c$ is the resulting velocity fraction after the boost and $\gamma = 1/\sqrt{1 - |\vec \beta|^2}$ is the Lorentz time-dilation factor. The basis of unit vectors for the four-momentum $\underline{p}$ can then be represented using the four Pauli matrices $(\hat{1},\hat{\sigma}_1,\hat{\sigma}_2,\hat{\sigma}_3)$, after which the Lorentz transformation has a double-sided matrix representation,
\begin{align}
    \hat p = mc \hat 1 \qquad&\longrightarrow \qquad \hat p' = \hat L \hat p \hat L^\dagger = \gamma mc \hat 1  + \gamma mc \sum_{k=1}^3 \beta_k \hat \sigma_k
\end{align}
involving the boost (spinor) matrix 
\begin{align}
    \hat L &= \exp(\alpha \hat v/2) = \cosh(\alpha/2)\hat{1} + \sinh(\alpha/2)\hat v , & \tanh\alpha &= |\vec \beta|, & \hat{v} &= \sum_{k=1}^3 \frac{v_k}{|\vec{v}|}\,\hat{\sigma}_k
\end{align}
that depends on the hyperbolic rapidity angle $\alpha$ and the matrix representation of the velocity unit vector direction $\hat v$. Using this matrix representation, the four-momentum $\hat{p}$ has the same structure as the unnormalized qubit density matrix $\hat{s}$ in Eq.~\eqref{eq:L_r}, while the boost transformation has the same structure as the qubit measurement in Eq.~\eqref{eq:measurement}. 

More generally, a Lorentz transformation matrix $\hat R = \hat{U}\hat{L} \in \text{SL}(2,\mathbb C)$ composed of both a unitary spatial rotation $\hat{U}$ and boost $\hat{L}$ is a two-sided (spinor) representation of a Lorentz transformation analogous to Eq.~\eqref{eq:R_r}. This equivalence arises because $\text{SL}(2,\mathbb C)$ is the double cover of the (proper orthochronous) Lorentz group $\text{SO}^+(1,3)$, in the same way that the sub-group of unitary qubit transformations $\text{SU}(2,\mathbb{R}) \subset \text{SL}(2,\mathbb{C})$ is a double cover of the orthogonal group $\text{SO}(3) \subset \text{SO}^+(1,3)$ of spatial rotations.  This representation of four-vectors and their Lorentz transformations has been long studied in the context of spin groups \cite{doranLieGroupsSpin1993,dreiner2010two,penroseSpinorsSpaceTimeVolume1984} and Clifford algebras \cite{doranStatesOperatorsSpacetime1993a, baylisComplexAlgebraPhysical2012}.

Pragmatically, this means that the 3-dimensional Bloch vector representation $\vec{S}$ (with transformations in $\text{SO}(3)$) of a normalized qubit state $\hat{\rho}$ (with transformations in $\text{SU}(2,\mathbb{R})$) can be directly generalized to a 4-dimensional vector in spacetime (with transformations in $\text{SO}^+(1,3)$) that represents an unnormalized qubit state $\hat{s}$ (with transformations in $\text{SL}(2,\mathbb{C})$). Conversely, a relativistic four-vector like the four-momentum $\underline{p}$ of a point particle can be represented as a matrix $\hat{p}$ analogous to an unnormalized qubit density matrix $\hat{s}$,
\begin{align}\label{eq:four-vector}
    \hat p &= \frac{E}{c} \hat 1 + p_x \hat \sigma_x + p_y \hat \sigma_y + p_z \hat \sigma_z = p^\mu \hat \sigma_\mu = p_\mu\hat{\sigma}^\mu, & \hat{s} &= s^\mu\hat{\sigma}_\mu = s_\mu \hat{\sigma}^\mu,
\end{align}
where for convenience we introduce contravariant (raised-index) components $p^\mu$ matched with the (lowered-index) Pauli basis $\hat{\sigma}_\mu$ and covariant (lowered-index) components $p_\mu = \eta_{\mu\nu}p^\nu$ matched with the (raised-index) reciprocal Pauli basis $\hat{\sigma}^\mu = \eta^{\mu\nu}\hat{\sigma}_{\nu}$, as well as the implied summation convention for matched pairs of raised and lowered Greek spacetime indices $\mu=0,1,2,3$. We use the spacetime metric $\eta_{\mu\nu}$ with signature $(+,-,-,-)$ and nonzero components $\eta_{00}=\eta^{00}=1$ and $\eta_{kk}=\eta^{kk}=-1$ for $k=1,2,3$.

The Minkowski inner product between four-vectors takes the matrix form 
\begin{align}
    \langle \hat a, \, \hat b \rangle &= \frac{1}{2} \Tr(\widetilde{\hat a} \, \hat b) = a_0 b_0 - \vec a \cdot \vec b = a^\mu b_\mu, & \langle \hat{\sigma}_\mu,\,\hat{\sigma}_\nu\rangle = \eta_{\mu\nu},
\end{align}
where the tilde denotes the Clifford conjugate for the Pauli algebra that transposes products, $\widetilde{\hat A \hat B} = \widetilde{\hat B} \widetilde{\hat A}$, and changes the Pauli basis to the reciprocal basis, e.g. $\widetilde{\hat p} = p_0 \hat 1 - p_i \hat \sigma_i$ \cite{doranGeometricAlgebraPhysicists2007, baylisComplexAlgebraPhysical2012,dreiner2010two}. The inner product is invariant under common Lorentz transformations, $\hat{a}\mapsto \hat{R}\hat{a}\hat{R}^\dagger$ and $\hat{b}\mapsto \hat{R}\hat{b}\hat{R}^\dagger$, because the conjugate of a Lorentz group element is its inverse, $\widetilde{\hat R} = \hat R^{-1}$. 

Notably, the determinant of the matrix $\hat p$ evaluates to its squared magnitude,
\begin{align}
    \det \hat p = \langle \hat p, \hat p \rangle = p_0^2 - \vec p \cdot \vec p = p^\mu p_\mu = (m c)^2,
\end{align}
and is invariant under Lorentz transformations since $\operatorname{det} (\hat R \hat p \hat R^\dagger) = \det \hat R \det p \det \hat R^\dagger = \det \hat p$.

The qubit analog of an invariant magnitude,
\begin{align}\label{eq:det-sigma}
    \det \hat s = \frac{\Tr(\hat s)^2 - \Tr(\hat s^2)}{2} = \Tr(\hat s)^2 \frac{1 - \Tr (\hat \rho^2)}{2} = s_0^2\, S_L(\hat \rho),
\end{align}
is proportional to the linear entropy $S_L(\hat \rho) = 2 (1 - \Tr(\hat \rho^2))$ satisfying $0 \leq S_L \leq 1$. Unlike $S_L(\hat \rho)$, $\det \hat s$ is invariant under measurement backaction. Similarly generalized definitions of linear entropy have been explored in the study of entropy production in non-Hermitian systems \cite{sergiLinearQuantumEntropy2016}. As such, pure states, which satisfy $\det\hat{s} = 0$, are the qubit analog of massless particles, which satisfy $\det\hat{p} = 0$. Similarly, mixed states, which satisfy $\det\hat{s} > 0$, are the qubit analog of massive particles, which satisfy $\det\hat{p} > 0$. Likewise, we find that the inner product between two unnormalized density matrices
\begin{align}
    \langle \hat a, \hat b \rangle = a_0 b_0 S_L(\hat \rho_A, \hat \rho_B)
\end{align}
is proportional to the linear cross entropy $S_L(\hat \rho_A, \hat \rho_B) = 2\left(1 - \Tr(\hat \rho_A \hat \rho_B)\right)$ of $\hat \rho_A = \hat a / \Tr \hat a$ and $\hat \rho_B = \hat b / \Tr \hat b$ \cite{shangnanQuantumCrossEntropy2022}.

\begin{figure}[]
    \centering
    {\includegraphics[width=\linewidth]{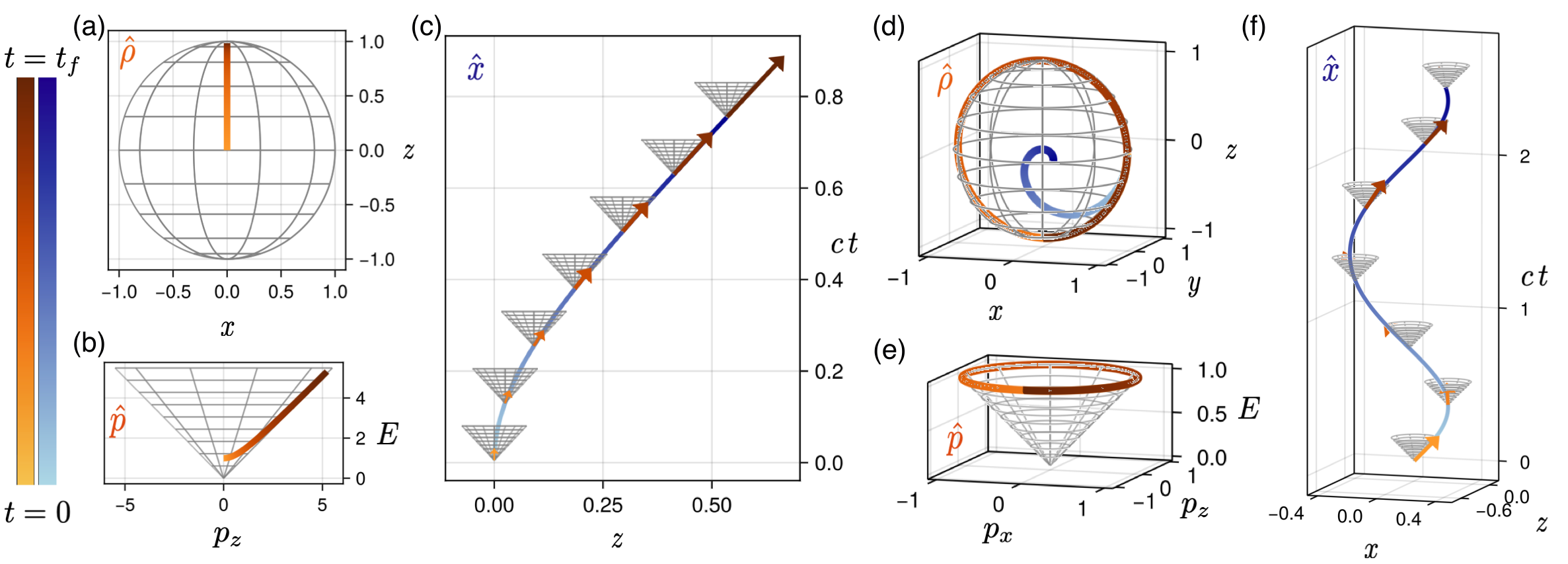}}
    \caption{Visualization of the correspondence between the structure of a qubit and a relativistic point particle for (a-c) hyperbolic purification of a maximally mixed state (corresponding to a massive particle at rest) and (d-f) rotational motion of a pure state $\ket{+x}$ (corresponding to a massless particle moving in the $x$ direction). Time evolution is represented in color going from light to dark. (a) The qubit state $\hat \rho$ (point particle velocity $\hat \beta$) is purified in the $z$ direction by measurement and asymptotically approaches a pure state (the speed of light). (b) The unnormalized qubit state $\hat s$ (particle momentum $\hat p$) is boosted in the $z$ direction, causing the momentum $p_z$ to asymptotically approach the surface of the lightcone. (c) The integrated velocity traces out a relativistic worldline $\hat x$ (cumulative state $\hat \phi$) in which the particle's lightcone is tangent at each point.  (d) The pure qubit state $\hat \rho$ (particle velocity $\hat \beta$) undergoes Rabi oscillations about $\hat \sigma_y$. (e) The unnormalized qubit state $\hat s$ corresponds to a particle with momentum $\hat p$ initially in the $x$ direction, accelerated around the $y$ axis. (f) The integrated velocity traces out a helical worldline $\hat x$ (cumulative state $\hat \phi$), which normalizes to the mean velocity $\overline{\hat \beta}$ (time averaged state $\overline{\hat \rho}$) in panel (d).
    }
\label{fig:projection}
\end{figure}

The spacetime analog of the normalized density matrix $\hat{\rho} = \hat{s}/\Tr(\hat{s})$ is the velocity fraction
\begin{align}\label{eq:beta}
    \hat \beta = \frac{\hat p}{\Tr \hat p} = \frac{1}{2}\left(\hat 1 + \frac{p_j}{p_0} \hat \sigma_j\right) = \frac{1}{2}\left(\hat 1 + \frac{v_j}{c} \hat \sigma_j\right)
\end{align}
due to the relation 
\begin{align}
    \frac{\vec p}{p_0} = \frac{\vec p c}{E} = \frac{\vec v}{c}
\end{align}
that holds for both massive and massless particles. In the special case of a massless particle, $\vec \beta = \vec v / c$ is the unit-vector direction of propagation. Massive particles at rest have a Bloch-like velocity description analogous to a maximally mixed qubit state, while massless particles have a velocity description analogous to pure qubit states,
\begin{align}
     \hat p = mc \hat 1 \mapsto \hat \beta = \frac{\hat p}{\Tr \hat p} = \frac{\hat 1}{2} \qquad&\qquad \hat p = \hbar(\omega \hat 1 / c + k_x \hat \sigma_x) \mapsto \hat \beta = \frac{\hat p}{\Tr(\hat p)} = \frac{1}{2}(\hat 1 + \hat \sigma_x) = \ket{x}\!\bra{x}.
\end{align}
The particle's spacetime displacement $\hat x$ is proportional to the time integral of the velocity fraction $\hat \beta$,
\begin{align}
    \hat x(t) &= 2c \int_0^t ds\, \hat \beta(s) = ct \hat 1 + \sum_{k=1}^3x_k(t) \hat \sigma_k & \hat \phi(t) &= 2 \int_0^t ds \,\hat \rho(s) = t \hat 1 + \sum_{k=1}^3 \int_0^t ds \,S_k(s) \hat \sigma_k,
\end{align}
while the cumulative qubit trajectory $\hat \phi(t)$ is the qubit analog of particle position and the factor of $2$ ensures $\langle \hat x/c, \hat \sigma_0 \rangle = \langle \hat \phi, \hat \sigma_0 \rangle = t$. Normalizing yields a Bloch-like description of the mean velocity fraction,
\begin{align}\label{eq:time-average}
    \overline{\hat \beta}(t) = \hat x(t)/\text{Tr} \, \hat x(t) = \frac{1}{t} \int_0^t ds \hat \beta(s) \qquad&\qquad \overline{\hat \rho}(t) = \hat \phi(t)/\text{Tr} \, \hat \phi(t) = \frac{1}{t} \int_0^t ds \hat \rho(s),
\end{align}
and its qubit analog $\overline{\hat \rho}(t)$: the time-averaged qubit state over its history. Fig.~\ref{fig:projection} depicts the quantities relevant to the kinematical analogy through examples of hyperbolic and rotational motion (introduced in Sec.~\ref{sec:review} and further elaborated in Sec.~\ref{sec:dynamics}), while
Table~\ref{tab:kinematical-correspondence} summarizes the formal correspondence between these quantities.

\begin{table}[]
    \def\arraystretch{1.75}
      \centering\small
    \begin{tabular}{ll|ll}
        \multicolumn{2}{c}{\textbf{Point Particle}} & \multicolumn{2}{c}{\textbf{Qubit}} \\ \hline \hline \vspace{-1em}\\
        \multicolumn{4}{c}{\textbf{Spacetime (Unnormalized) Description}} \\ \hline
        Four-momentum & Lorentz transformation & Unnormalized state & Evolution and measurement \\
        $\quad\hat p = p_0 \hat 1 + \sum_{k=1}^3 p^k \hat \sigma_k$ & $\quad\hat p \mapsto \hat R \hat p \hat R^\dagger$ & $\quad \hat s = s_0 \hat 1 + \sum_{k=1}^3 s^k \hat \sigma_k$ & $\quad\hat s \mapsto \hat R \hat s \hat R^\dagger$ \\ \hline
        Invariant mass & Minkowski inner product & Linear entropy & Linear cross entropy\\
        $\quad \det \hat p = (mc)^2$ & $\langle \hat p_1, \hat p_2 \rangle = E_1E_2/c^2 - \vec p_1 \cdot \vec p_2$ & $\quad \det \hat s = s_0^2 S_L(\hat \rho)$ & $\langle \hat a, \hat b \rangle = a_0 b_0 S_L(\hat \rho_A, \hat \rho_B)$\\ \hline
        Massless particle & Propagating along $x$ & Pure state & Polarized along $x$ \\
        $\quad\det \hat p = 0$ & $\quad\hat p = E(\hat 1 + \hat \sigma_x)/c$ & $\quad\det \hat s = 0$ & $\quad\hat s = s_0 (\hat 1 + \hat \sigma_x)$  \\ \hline
        Massive particle & At rest & Mixed state & Maximally mixed \\
        $\quad\det \hat p > 0$ & $\quad\hat p = mc \hat 1$ & $\quad\det \hat s > 0$ & $\quad\hat s = s_0 \hat 1$ \\ \hline
        Displacement  && Cumulative state & \\
        \multicolumn{2}{l|}{$\quad\hat x(t) = 2c \int_0^t ds \hat \beta(s)$} & \multicolumn{2}{l}{$\quad\hat \phi = 2 \int_0^t ds \hat \rho(s) $} \\  \\
        \multicolumn{4}{c}{\textbf{Bloch Ball (Normalized) Description}} \\ \hline
        Velocity fraction && Normalized state  \\
        \multicolumn{2}{l|}{$\quad\hat \beta = \hat p/\Tr \hat p = [\hat 1 + \sum_{k=1}^3 (v_k/c) \, \hat \sigma_k]/2$} & \multicolumn{2}{l}{$\quad\hat \rho = \hat s/\Tr \hat s = [\hat 1 + \sum_{k=1}^3 S_k \hat\sigma_k]/2$} \\ \hline
        Massless particle & Propagating along $x$ & Pure state & Polarized along $x$ \\
        $\quad\Tr\hat \beta^2 = 1$ & $\quad \hat \beta = (\hat 1 + \hat \sigma_x)/2$ & $\quad\Tr\hat \rho^2 = 1$ & $\quad \hat \rho = (\hat 1 + \hat \sigma_x)/2$ \\ \hline
        Massive particle & Particle at rest & Mixed state & Maximally mixed \\
        $\quad\Tr\hat \beta^2 < 1$ & $\quad\hat \beta = \hat 1/2$ & $\quad\Tr\hat \rho^2 < 1$ & $\quad \hat \rho = \hat 1/2$ \\ \hline
        Mean velocity && Mean state trajectory \\
        \multicolumn{2}{l|}{$\quad \overline{\hat \beta} = \hat x/\Tr \hat x = \int_0^t ds \hat \beta(s)/t$} & \multicolumn{2}{l}{$\quad \overline{\hat \rho} = \hat \phi/\Tr \hat \phi = \int_0^t ds \hat \rho(s)/t$} 
    \end{tabular}
    \caption{Summary of the kinematical correspondence between a relativistic point particle (left) and qubit (right). Unnormalized quantities exist in a 4-dimensional spacetime, while normalized quantities exist in a 3-dimensional Bloch ball. }\label{tab:kinematical-correspondence}
\end{table}

\section{Dynamical Correspondence} \label{sec:dynamics}

We now draw a formal correspondence between the deterministic dynamics of a point charge momentum $\hat{p}$ and the deterministic dynamics of an unnormalized qubit state $\hat{s}$. (For clarity, we delay treatment of stochastic dynamics until Section~\ref{sec:stochastics}.) Magnetic fields are the generators of spatial rotations as part of the Lorentz force, and correspond to the unitary Hamiltonian evolution of a qubit. Electric fields are the generators of boosts causing linear acceleration as part of the Lorentz force, and correspond to deterministic measurement backaction on a qubit. Such deterministic non-unitary dynamics generated by a non-Hermitian Hamiltonian has been used productively to model conditional dissipation in open qubit systems, such as that obtained by post-selection on particular outcomes of a dissipative interaction with a bath or measurement apparatus \cite{carmichaelOpenSystemsApproach1993, plenioQuantumjumpApproachDissipative1998, naghilooQuantumStateTomography2019, chenQuantumJumpsNonHermitian2021, roccatiNonHermitianPhysicsMaster2022, martinez-azconaQuantumDynamicsStochastic2025}.

Focusing first on unitary spatial rotations, the evolution for both a point charge momentum $\hat p$ and an unnormalized qubit state $\hat s$ with a time parameter $t$ is generated by an anti-Hermitian operator $-i\hat H = i\mu \hat{B}$ with units of energy and components that we parameterize with an effective magnetic field $\hat{B} = \sum_{k=1}^3 B_k \hat \sigma_k$ up to a scaling factor $\mu$,
\begin{align}\label{eq:magnetic-dynamics}
    \hat p(t + dt) = e^{i \mu dt \hat B/\hbar} \hat p(t) e^{-i \mu dt \hat B/\hbar} \qquad\qquad \hat s(t + dt) = e^{-idt \hat H/\hbar} \hat s(t) e^{idt \hat H/\hbar}.
\end{align}
We formally include a factor of $\hbar$ to make the exponents properly dimensionless in both cases, and to follow quantum mechanical convention.
Expanding to first order in $dt$ yields corresponding von Neumann differential equations that are well-known from Schrödinger-picture quantum dynamics,
\begin{align}
    \frac{d \hat p}{dt} = \frac{i\mu}{\hbar} [\hat B, \hat p] \qquad\qquad \frac{d \hat s}{dt} = -\frac{i}{\hbar} [\hat H, \hat s],
\end{align}
where $[\hat{A},\hat{B}] = \hat{A}\hat{B} - \hat{B}\hat{A}$ is a commutator. Expressing these equations in familiar 3-vector notation using Eq.~\eqref{eq:four-vector} and choosing $\mu = \mu_B (mc/p_0) = (e\hbar/2m)(mc/p_0) = e\hbar c/2p_0$ to be the relativistic generalization of the Bohr magneton (recalling that $\vec{p}/p_0 = \vec{v}/c$), yields
\begin{align}\label{eq:rotational}
    c\frac{dp_0}{dt} &= W_B = 0, \qquad \frac{d \vec p}{dt} = \vec{F}_B = e \vec v \times \vec B \qquad\qquad \frac{d s_0}{dt} = 0, \qquad \frac{d\vec s}{dt} =  \frac{2\mu_B}{\hbar} \vec s \times \vec B.
\end{align}
We thus recover the expected expressions of magnetic work $W_B$ and magnetic force $\vec{F}_B$ for a massive point charge. Notably, since $mc/p_0 = d\tau/dt$, where $\tau$ is the proper time for a massive point particle, the scaled Bohr magneton $\mu = \mu_B\,d\tau/dt$ is actually momentum-independent and implicitly changes coordinates back to the proper time $\tau$ inside the exponent of the transformation in Eq.~\eqref{eq:magnetic-dynamics} as a distinguished parameterization for the particle path. For a massless charge with momentum $\underline{p} = (p_0,\vec{p}) = (\hbar\omega/c,\hbar\vec{k})$ such that $|\vec{k}|=\omega/c$, there is no distinguished rest frame with proper time parameter $\tau$, but the choice $\mu = e\hbar c/2p_0 = ec^2/2\omega$ will still remain correctly matched to the time parameter $t$ on a change of frame and yield the same equation $d\vec{p}/dt = e\,\vec{v}\times\vec{B}$ with $\vec{p}/p_0 = \vec{v}/c = \vec{\beta}$ and $|\vec{\beta}| = 1$. Similarly, after restricting to the non-relativistic (rest frame) Bohr magneton $\mu = \mu_B$, we also recover the expected non-relativistic spin-1/2 Hamiltonian $\hat{H} = -\mu_B\sum_{k=1}^3 B_k\hat{\sigma}_k$ for the energy of a magnetic dipole as well as the expected spin-1/2 state precession dynamics from an external magnetic field.

The probability preserving property $ds_0/dt = 0$ of Hamiltonian evolution for a qubit is analogous to energy conservation $dp_0/dt = 0$ for a point charge due to the fact that magnetic fields do no work. This conservation ensures that the dynamics of the corresponding normalized quantities $\hat \beta = \hat p / \Tr \hat p$ and $\hat \rho = \hat s / \Tr \hat s$ have the same von Neumann form,
\begin{align}
    \frac{d \hat \beta}{dt} &= i \frac{\mu}{\hbar} [\hat B, \hat \beta] \; \mapsto \; \frac{d\vec{v}}{dt} = \vec{a} = \frac{ec}{p_0}\,\vec{v}\times\vec{B}, & \frac{d \hat \rho}{d t} &= -\frac{i}{\hbar}[\hat H, \hat \rho] \; \mapsto \; \frac{d\vec{S}}{dt} = \frac{2\mu_B}{\hbar}\vec{S}\times\vec{B},
\end{align}
recovering the expected magnetic acceleration $\vec{a}$ of a point charge as well as the expected precession dynamics for the qubit Bloch spin-vector $\vec{S} = \vec{s}/s_0$.

Focusing next on nonunitary boost accelerations, the evolution for both a point charge momentum $\hat{p}$ and an unnormalized qubit state $\hat{s}$ is generated by a Hermitian operator $\hat D = \mu \hat{E}/c$ with units of energy and components that we similarly parameterize with an effective electric field $\hat{E}= \sum_{k=1}^3 E_k \hat \sigma_k$ up to a scaling factor $\mu$, 
\begin{align}
    \hat p(t + dt) = e^{\mu dt \hat E/c\hbar} \hat p(t) e^{\mu dt \hat E/c\hbar} \qquad&\qquad \hat s(t + dt) = e^{dt \hat D/\hbar} \hat s(t) e^{dt \hat D/\hbar}.
\end{align}
Expanding to first order in $dt$ yields differential equations,
\begin{align}
    \frac{d \hat p}{dt} &= \frac{\mu}{\hbar c} \{ \hat E,  \hat p \}, & \frac{d \hat s}{dt} &= \frac{1}{\hbar} \{ \hat D, \hat s \},
\end{align}
that involve an anti-commutator $\{\hat{A},\hat{B}\} = \hat{A}\hat{B} + \hat{B}\hat{A}$ instead of a commutator. Choosing again the relativistic Bohr magneton $\mu = \mu_B (mc/p_0) = e\hbar c/2p_0$ and expanding into 3-vector notation yields,
\begin{align}
    c \frac{dp_0}{dt} = W_E = \vec v \cdot (e\vec E), \qquad \frac{d \vec p}{dt} = \vec{F}_E = e \vec E, \qquad\qquad &\qquad \frac{d s_0}{dt} = \frac{2\mu_B}{\hbar c} \vec s \cdot\vec E, \qquad \frac{d\vec s}{dt} = \frac{2\mu_B}{\hbar c} s_0 \vec E.
\end{align}
We thus recover the expected expressions of electric work $W_E$ and electric force $\vec{F}_E$ for a massive point charge. As with the magnetic case in Eq.~\eqref{eq:magnetic-dynamics}, the scaling factor $\mu = \mu_B(d\tau/dt)$ implicitly changes coordinates back to the proper time $\tau$ inside the exponent of the transformation for the massive case, while for the massless case the factor becomes $\mu = e\hbar c/2p_0 = ec^2/2\omega$ and is explicitly matched with the chosen $t$ parameter to remain invariant under changes of frame. Similarly, choosing the nonrelativistic $\mu = \mu_B$ in the qubit case produces collapse-like dynamics for spin-1/2 generated by an effectively non-Hermitian Hamiltonian $\hat{H}_E = i\hat{D} = i\mu_B\hat{E}/c$.

Electric fields generally do work, so do not conserve energy, $dp_0/dt\neq 0$. Similarly, non-Hermitian qubit dynamics generally do not conserve probability, so $ds_0/dt\neq 0$. This non-conservation induces nonlinear dynamics for the corresponding normalized quantities $\hat \beta = \hat p / \Tr \hat p$ and $\hat \rho = \hat s / \Tr \hat s$,
\begin{align}
    \frac{d \hat \beta}{dt} &= \frac{\mu}{\hbar c} \{\hat E, \hat \beta\} - \frac{2\mu}{\hbar c} \operatorname{Tr}(\hat E \hat \beta) \hat \beta, &
    \frac{d \hat \rho}{dt} &= \frac{1}{\hbar} \{ \hat D,  \hat \rho \} - \frac{2}{\hbar} \operatorname{Tr}(\hat D \hat \rho) \hat \rho.
\end{align}
Expanding these equations in 3-vector form using $\mu = e\hbar c/2p_0$ yields,
\begin{align}\label{eq:boost-normalized}
    c\frac{d \vec \beta}{dt} &= \vec{a} = \frac{ec}{p_0} \left[\vec E - (\vec \beta \cdot \vec E) \vec \beta\right], &
    \frac{d \vec S}{dt} &= \frac{2\mu_B}{\hbar c}\left[\vec E - 2 (\vec S \cdot \vec E)\vec S\right],
\end{align}
which recovers the correct electric acceleration for a point charge, where the role of the nonlinear correction term is to ensure causal propagation with $|\vec \beta| \leq 1$. For a normalized qubit state, the corresponding nonlinear correction renormalizes the state to continually reset the total probability back to 1 such that the normalized Bloch vector satisfies $|\vec{S}|\leq 1$.

We now combine the two cases above to consider the action of a full Lorentz transformation involving both spatial rotation and boost acceleration. The corresponding generator $\hat G = \hat D - i \hat H = \mu (\hat E/c + i \hat B) = \mu \hat F$ includes both the Hermitian boost generator $\hat{D}$ and anti-Hermitian rotation generator $-i\hat{H}$ and thus is parameterized by the full electromagnetic field $\hat{F}$ (in Riemann-Silberstein form \cite{dressel_spacetime_2015}), 
\begin{align}\label{eq:non-hermitian}
    \hat p(t + dt) &= e^{\mu dt \hat F/\hbar} \hat p(t) e^{\mu dt \hat F^\dagger/\hbar}, &
    \hat s(t + dt) &= e^{dt \hat G/\hbar} \hat s(t) e^{dt \hat G^\dagger/\hbar}.
\end{align}
Expanding to first order in $dt$ yields evolution familiar from deterministic non-Hermitian qubit dynamics,
\begin{align}\label{eq:lorentz-matrix}
    \frac{d\hat p}{dt} &= \frac{2\mu}{\hbar} \big(\hat F \hat p + \hat p \hat F^\dagger \big) = \frac{2\mu}{\hbar} \{ \hat F, \hat p \}, &
    \frac{d \hat s}{dt} &= -\frac{i}{\hbar} \big(\hat H_\text{eff} \hat s - \hat s \hat H_\text{eff}^\dagger\big) = -\frac{i}{\hbar} [\hat H_\text{eff}, \hat s ],
\end{align}
in terms of non-Hermitian generalizations of the anti-commutator and commutator,
\begin{align}
    \{ \hat A, \hat B \} &= \hat A \hat B + \hat B \hat A^\dagger, &
    [\hat A, \hat B] &= \hat A \hat B - \hat B \hat A^\dagger,
\end{align}
and where $\hat H_\text{eff} = i \hat G = \hat H + i \hat D = i \mu \hat F$ is the effective non-Hermitian Hamiltonian. Expanding Eqs.~\eqref{eq:lorentz-matrix} in 3-vector form using $\mu = e\hbar c/2p_0$ yields the expected work $W$ and Lorentz force $\vec{F}_L$ for a point charge:
\begin{align}\label{eq:lorentz-vector}
    c\frac{dp_0}{dt} &= W = \vec v \cdot (e \vec E), \quad \frac{d \vec p}{dt} = \vec{F}_L = e \vec E + e\vec v \times \vec B, \quad&
    \frac{ds_0}{dt} &= \frac{2\mu_B}{\hbar c} \vec s \cdot \vec E, \quad \frac{d\vec s}{dt} = \frac{2 \mu_B}{\hbar c}\left( s_0 \vec E + c\vec s \times \vec B\right).
\end{align}
Similarly, using the non-relativistic $\mu = \mu_B$ in the qubit case produces the spin-1/2 dynamics expected from a general non-Hermitian Hamiltonian.

The corresponding dynamics of the normalized states $\hat \beta = \hat p / \Tr \hat p$ and $\hat \rho = \hat s / \Tr \hat s$ become nonlinear in the state due to the continuous renormalization \cite{gisinSimpleNonlinearDissipative1981, goetschLinearStochasticWave1994, brodyMixedStateEvolutionPresence2012, wakefieldNonHermiticityQuantumNonlinear2024},
\begin{align}\label{eq:lorentz-normalized-matrix}
    \frac{d \hat \beta}{dt} &= \frac{\mu}{\hbar} \left(\{ \hat F, \hat \beta \} - \Tr\left[(\hat F + \hat F^\dagger) \hat \beta\right] \hat \beta\right), &
    \frac{d \hat \rho}{dt} &= -\frac{i}{\hbar} \left([\hat H_\text{eff}, \hat \rho] - \Tr\left[(\hat H_\text{eff} - \hat H_\text{eff}^\dagger)\hat \rho \right] \hat \rho\right).
\end{align}
Expanding these equations into 3-vector form using $\mu = e\hbar c/2 p_0$ yields,
\begin{align}\label{eq:lorentz-normalized-vector}
    c\frac{d \vec \beta}{dt} &= \vec{a} = \frac{e c}{p_0} \left(\vec E + \vec \beta \times \vec B - (\vec E \cdot \vec \beta) \vec \beta\right), &
    \frac{d \vec S}{dt} = \frac{2\mu_B}{\hbar c} \left(\vec E + c\vec S \times \vec B - (\vec E \cdot \vec S) \vec S\right),
\end{align}
thus recovering the expected Lorentz acceleration for a point charge. In the qubit case using $\mu = \mu_B$ recovers the expected nonlinear dynamics of the normalized spin-1/2 Bloch vector $|\vec{S}|\leq 1$ under deterministic non-Hermitian dynamics. Note that demanding linearity in $\hat \beta$ and $\hat \rho$ forces $\vec E$ and $\hat D$ to vanish and restricts to standard Hermitian quantum theory.

\begin{figure}[]
    \centering
    \includegraphics[width=\linewidth]{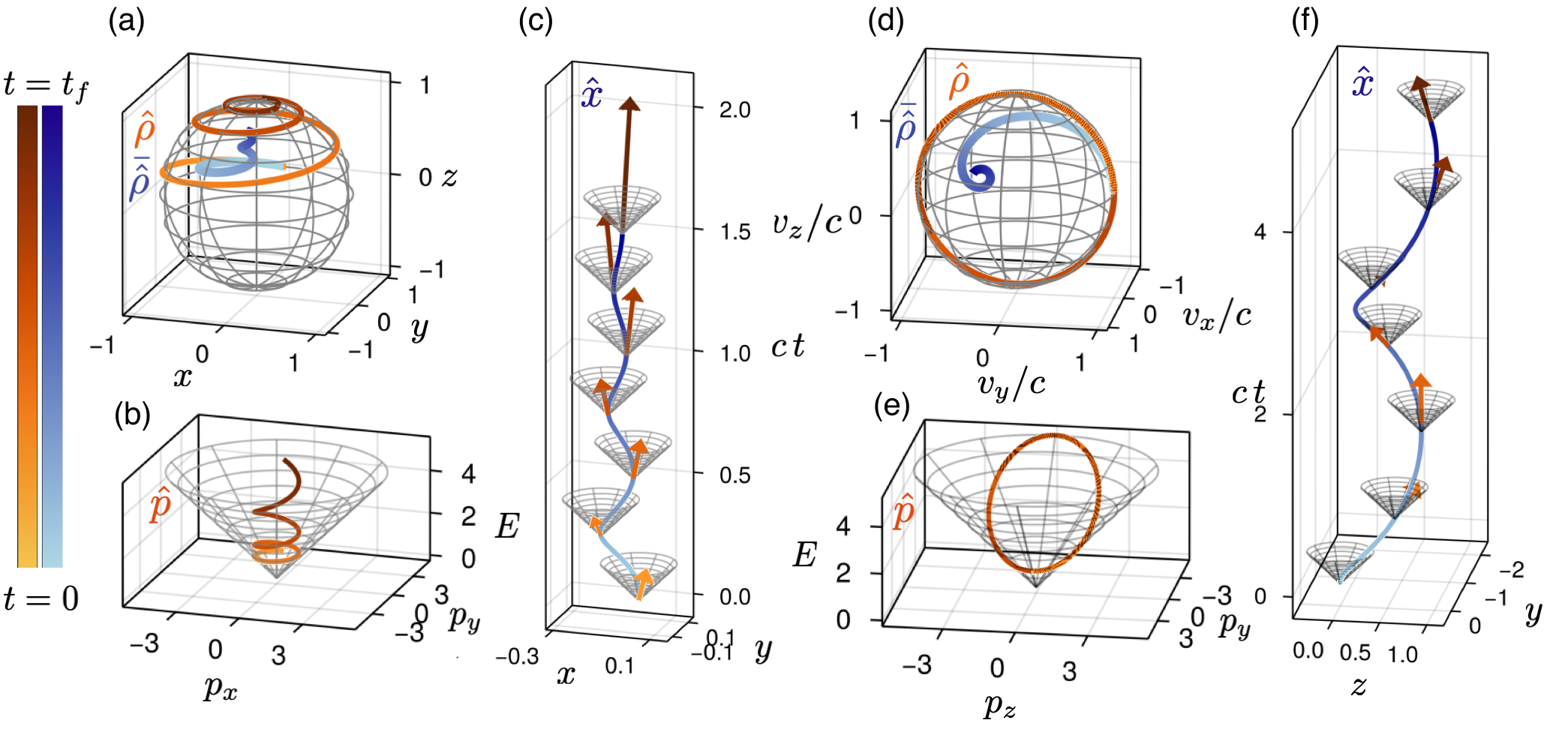}
    \caption{
    Combined hyperbolic and rotational motions for a pure qubit state initially in $\ket{+y}$, corresponding to a massless particle with velocity initially moving along $+y$. (a-c) Evolution generated by $\hat H_\mathrm{eff} = \mu (B_z + iE_z/c)\hat{\sigma}_z$ with $E_z / c B_z = 1/8$. (a) The non-Hermitian boost in $\hat \sigma_z$ generates hyperbolic evolution of $\hat \rho$ ($\hat \beta$) towards the qubit $\ket 0$ state ($z$ velocity), with simultaneous rotation about $\hat \sigma_z$ resulting in spiraling motion that stabilizes at the fixed point $\ket 0$. (b-c) The helical momentum $\hat p$ (unnormalized qubit state $\hat s$) (b) and worldline $\hat x$ (c) appear increasingly timelike in the $x$-$y$ slice of spacetime as the particle is boosted in the $z$ direction, corresponding to the normalized qubit state $\hat \phi$ (blue, panel (a)) converging to the fixed point. (d-f) Evolution  generated  by $\hat H_\mathrm{eff} = \mu (B_x\hat{\sigma}_x + i(E_z/c)\hat{\sigma}_z)$ with $E_z / c B_x = 2/3$. (d) The non-commutativity of $\hat \sigma_x$ and $\hat \sigma_z$, with $E_z /c B_x < 1$, gives rise to Bloch sphere rotations with modulated effective frequency, evidenced by the bias of the mean velocity (blue) towards $\ket{-y}$. (f) Elliptical helix of the worldline, elongated due to the competing electric and magnetic forces. 
    }
    \label{fig:lorentz}
\end{figure}

The interplay between hyperbolic and rotational dynamics in electromagnetism provides useful intuition for the analogous non-Hermitian qubit dynamics. Fig.~\ref{fig:lorentz}(a-c) depicts the simple example of a massless charge in constant electric and magnetic fields---analogously, a pure qubit state undergoing simultaneous hyperbolic and rotational motion. In this example, the electric field $\hat E = E_z \hat \sigma_z$ and magnetic field $\hat B = B_z \hat \sigma_z$ commute, paralleling the structure introduced for qubit measurements in Sec.~\ref{sec:review}. This results in a helical velocity trajectory $\hat \beta(t)$ on the Bloch sphere converging towards a fixed point, sweeping out a helical worldline $\hat x(t)$ that tightens over time as the particle is accelerated in the $z$ direction and a helical momentum trajectory $\hat p(t)$ in the $p_x$-$p_y$ slice of the lightcone. Notably, the helical path on the Bloch sphere is determined entirely by the angle $\arctan(E_z/cB_z)$ such that $\hat H_\mathrm{eff} \propto e^{i \arctan(E_z/cB_z)} \hat \sigma_z$, which rotates the Hermitian operator $\hat \sigma_z$ smoothly into the anti-Hermitian operator $i \hat \sigma_z$, offering intuition for the behavior of qubit stochastics on the Bloch sphere that we will elaborate on in Sec.~\ref{sec:delay}. 

Interestingly, for the special case of \emph{massless} point charges with $|\vec \beta| = 1$, Eqs.~\eqref{eq:lorentz-normalized-vector} can be written in an equivalent form that appears entirely magnetic. That is, using the vector identity $\vec a \times (\vec b \times \vec a) = (\vec a \cdot \vec a) \vec b - (\vec a \cdot \vec b) \vec a$ and $\vec \beta \cdot \vec \beta = 1$, Eq.~\eqref{eq:lorentz-normalized-vector} is equivalent to \cite{brodyMixedStateEvolutionPresence2012}
\begin{align}
    c\frac{d \vec \beta}{dt} &= \vec{a} = \frac{e c}{p_0} \vec \beta \times (\vec B + \vec B_E), &
    \vec{B}_E &= -\vec{\beta}\times\vec{E}, & 
    |\vec{\beta}| = 1,
\end{align}
which looks like a purely magnetic acceleration from an effective \emph{velocity-dependent} magnetic field $\vec{B}_E$ that emulates the boost effect of the equivalent electric field $\vec{E}$. The physical reason for this equivalence is that the speed of a massless particle must always be $c$, so only the direction of the velocity can change. Thus, there always exists an effective magnetic field that can equivalently cause such a change in direction. In contrast to the velocity, the momentum evolution can clearly distinguish boosts from rotations due to the induced changes in energy $E = cp_0$. 

Similarly, for the special case of a \emph{pure} qubit state with $|\vec{S}| = 1$, Eqs.~\eqref{eq:lorentz-normalized-matrix} can be written in an equivalent form that appears generated by a Hermitian Hamiltonian. That is, for Hermitian $\hat{A}$ and $\hat{B}$ such that $\hat{A}^2 = \hat{A}$ and $\hat{A}\hat{B}\hat{A} = \operatorname{Tr}(\hat{A}\hat{B})\hat{A}$ the commutator identity $[\hat{A},[\hat{A},\hat{B}]] = \{\hat{A}^2,\hat{B}\} - 2\hat{A}\hat{B}\hat{A}$ simplifies to $\{\hat{A},\hat{B}\} - 2\operatorname{Tr}(\hat{A}\hat{B})\hat{A}$, so Eq.~\eqref{eq:lorentz-normalized-matrix} is equivalent to
\begin{align}\label{eq:false-unitary}
    \frac{d \hat \rho}{dt} &= -\frac{i}{\hbar} [\hat H + \hat H_D, \hat \rho], &
    \hat H_D &= -i[\hat \rho, \hat D], & 
    \hat{\rho}^2 = \hat{\rho},
\end{align}
which looks like purely Hermitian Hamiltonian dynamics from an effective \emph{state-dependent} (i.e., \emph{feedback}) Hamiltonian $\hat{H}_D$ that emulates the collapse-like effects of the equivalent non-Hermitian Hamiltonian $i\hat{D}$. The reason for this equivalence is that a pure state represents maximum certainty and is determined by its orientation relative to other states of maximum certainty, encoded as the direction of its unit Bloch vector $\vec{S}$. Thus, there always exists a Hermitian Hamiltonian that can equivalently cause such a Bloch vector rotation. In contrast, the evolution for the unnormalized state $\hat{s}$ can clearly distinguish Hermitian from non-Hermitian dynamics due to induced changes to the state normalization $s_0$, which will be proportional to the total probability of the observed evolution. 

Fig.~\ref{fig:lorentz}(d-f) displays a second example of simultaneous hyperbolic and rotational dynamics in which the electric field $\hat E = E_z \hat \sigma_z$ and the magnetic field $\hat B = B_x \hat \sigma_x$ anti-commute. This gives rise to purely rotational motion with a modulated frequency of revolution around a great circle of the Bloch sphere, sweeping out an elliptical helix in the $y$-$z$ slice of spacetime and a tilted, elliptical momentum trajectory on its lightcone. As previously alluded to, this pure, normalized state trajectory can be generated either by the non-Hermitian Hamiltonian $\hat{H} + i\hat{D} = \mu_B (-B_x\hat{\sigma}_x + i(E_z/c)\hat{\sigma}_z)$ that includes a deterministic drift towards negative $y$, or by a time-dependent, Hermitian feedback-control Hamiltonian $\hat{H} + \hat{H}_D(t)$ with $\hat{H}_D(t) = (\mu_B E_z/c)\, [\hat \rho, \hat \sigma_z]$ using knowledge of the state to modulate its rate of rotation around the $x$-axis. However, while the pure normalized state dynamics are equivalently generated by a non-Hermitian or a Hermitian Hamiltonian, which correspond to two very different implementations for a physical qubit, the unnormalized state dynamics distinguish the two cases. While a feedback control Hamiltonian could reproduce the same normalized dynamics as Fig.~\ref{fig:lorentz}(d), the momentum trajectory generated under this Hamiltonian would be flat due to probability conservation, rather than tilted as in Fig.~\ref{fig:lorentz}(e), due to the probability non-conservation of non-Hermitian evolution (equivalently, the energy non-conservation of boosts). Thus, the unnormalized state evolution encodes the Hermiticity of the generating Hamiltonian even for pure states, a notable feature that is absent in the normalized state.

\begin{figure}
    \vspace{2em}
    \def\arraystretch{1.75}
      \centering\small
    \begin{tabular}{l|l}
        \multicolumn{1}{c}{\textbf{Point Charge}} & \multicolumn{1}{c}{\textbf{Qubit}} \\ \hline \hline \multicolumn{2}{c}{}\vspace{-1em}\\
        \multicolumn{2}{c}{\textbf{Spacetime (Unnormalized) Dynamics}} \\ \hline
        \multicolumn{2}{c}{Generators of Lorentz Transformations} \\
        Electromagnetic Field & Effective (Non-Hermitian) Hamiltonian \\
        $\quad\hat F = \hat E/c + i \hat B$ & $\quad- i\hat H_\text{eff} = -i(\hat H + i \hat D)$ \\ 
        \hline
        \multicolumn{2}{c}{Lorentz Transformations} \\
        $\quad\begin{aligned}
            \hat p(t + dt) &= e^{\mu dt\,\hat F/\hbar} \hat p(t) e^{\mu dt\, \hat F^\dagger/\hbar}
        \end{aligned}$ 
        & 
        $\quad \begin{aligned}\hat s(t + dt) &= e^{-i dt\,\hat H_\text{eff}/\hbar} \hat s(t) e^{i dt\,\hat H_\text{eff}^\dagger/\hbar}\end{aligned}$ \\
        \hline
        Lorentz Force & von Neumann Equation \\
        $\quad d\hat p/dt =  \mu \{\hat F, \hat p\}/\hbar, \qquad \mu = \mu_B \, d\tau/dt$ & $\quad d \hat s/dt = -i [\hat H_\text{eff}, \hat s]/\hbar$ \\
        $\quad \mu_B = (e \hbar)/2m, \qquad d\tau/dt = mc/p_0$ & \\
        \multicolumn{2}{c}{}\vspace{-1em}\\
        \multicolumn{2}{c}{\textbf{Bloch Ball (Normalized) Dynamics}} \\ \hline
        Velocity trajectory & Qubit trajectory \\
        $\quad\hat \beta = \displaystyle\frac{1}{2}\bigg(\hat 1 + \sum_{k=1}^3 \beta_k \hat \sigma_k \bigg)$ & $\quad\hat \rho = \displaystyle\frac{1}{2}\bigg(\hat 1 + \sum_{k=1}^3 S_k \hat \sigma_k \bigg)$ \\
        $\text{Velocity ratio } \vec \beta = \vec p/p_0 = \vec v/c$ & $\text{Bloch vector } \vec S = \vec s / s_0$ \\
        \hline
        Normalized Lorentz force & Normalized von Neumann equation \\
        $\quad d \vec \beta/dt = 2\mu\left[\vec E + (c\vec \beta) \times \vec B - (\vec E \cdot \vec \beta) \vec \beta \right]/\hbar$ & $\quad d \vec S/dt = 2\left[\vec D + \vec S \times \vec H - (\vec D \cdot \vec S) \vec S\right]/\hbar$ 
    \end{tabular}

    \captionof{table}{Summary of the dynamical correspondence between a relativistic point charge (left) and a qubit (right). Unnormalized quantities exist in a 4-dimensional spacetime, while normalized quantities exist in a 3-dimensional Bloch ball. The commutator and anticommutator are defined as $[\hat A, \hat B] := \hat A \hat B - \hat B \hat A^\dagger$ and $\{ \hat A, \hat B\} = \hat A \hat B + \hat B \hat A^\dagger$, respectively.}\label{tab:dynamical-correspondence}
\end{figure}

Non-Hermitian Hamiltonians of the form $\hat H_\text{eff} =  \mu_B (-B_x\hat{\sigma}_x + i(E_z/c)\hat{\sigma}_z)$ as in Fig.~\ref{fig:lorentz}(d-f) are considered in the analysis of dissipative qubit dynamics near exceptional points \cite{murchObservingSingleQuantum2013a}, which correspond here to the case $\operatorname{Tr}(\hat F^2)/2 = |\vec E|^2/c^2 - |\hat B|^2 = 0$. More generally, the Lorentz invariants of the electromagnetic field classify the motion of a test charge suspended in the fields, ranging between elliptical motion $\operatorname{Tr}(\hat F^2) < 0$ (i.e. PT-symmetric phase of a dissipative qubit), parabolic $\operatorname{Tr}(\hat F^2) = 0$ (i.e. exceptional phase), and hyperbolic motion $\operatorname{Tr}(\hat F^2) > 0$ (i.e. PT-broken phase).

\section{Stochastic Correspondence} \label{sec:stochastics}

The previous section established a formal correspondence between the deterministic Lorentz-force dynamics of a point charge in spacetime and the deterministic dynamics of a qubit subject to a non-Hermitian Hamiltonian. However, as reviewed in Section~\ref{sec:review}, qubit measurement dynamics are generally stochastic, so should correspond to the dynamics of a point charge subject to stochastic Lorentz forces caused by fluctuating electromagnetic fields. In this section, we examine the dynamical structure of such stochastic Lorentz forces and highlight the peculiar constraints the fluctuating electromagnetic fields must satisfy to fully correspond to stochastic qubit measurement dynamics. 

To accommodate stochastic Lorentz transformations, 
the deterministic generator in Eq.~\eqref{eq:non-hermitian} generalizes to a stochastic process, and the dynamical map that advances by this time increment has the same form as Eq.~\eqref{eq:non-hermitian},
\begin{align}\label{eq:stochastic-lorentz-transformation}
    \hat p(t + dt) = e^{\mu dt \hat F/\hbar} \hat p(t) e^{\mu dt \hat F^\dagger/\hbar} \qquad&\qquad \hat s(t + dt) = e^{-i dt \hat H_\text{eff}/\hbar} \hat s(t) e^{i dt \hat H_\text{eff}^\dagger/\hbar}.
\end{align}
This group structure is thus independent of the specifics of the fluctuations that are implicit to $dt \hat{F}$.

Anticipating a correspondence to the diffusive qubit measurement in Eq.~\eqref{eq:stochastic-me}, we focus on fluctuations that can be approximated by additive zero-mean white noise. In this case, the stochastic generator
\begin{align}\label{eq:stochastic-non-hermitian-generator}
    \hat F &= \hat F_D + \hat F_\xi, & 
    \hat{H}_\text{eff} &= \hat H_D + \hat H_\xi
\end{align}
can be separated into deterministic generators of the mean dynamics $\mu dt\hat F_D = -idt\hat H_D$ that scale linearly with $dt$, and fluctuating zero-mean generators $\mu dt  \hat F_\xi = -i dt \hat H_\xi$ that scale as $\sqrt{dt}$ such that $(\mu dt \hat F_\xi)^2 \propto dt$ and $(-i dt \hat H_\xi)^2 \propto dt$. The fluctuations $dt \hat F_\xi$ may be decomposed into real and imaginary parts $dt \hat{F}_\xi = dt \hat{E}_\xi/c + i dt \hat{B}_\xi$ that correspond to the electric and magnetic field components of the fluctuations, which may or may not be correlated. The ensemble covariance properties of the fluctuations are characterized by their quadratic combinations: (i) the (frame-dependent) four-vector to first order in $dt$,
\begin{align}\label{eq:noise-poynting}
    (dt \hat F_\xi)^\dagger (dt \hat F_\xi) &= \left(\frac{dt \vec E_\xi \cdot dt \vec E_\xi}{c^2} + dt \vec B_\xi \cdot d t\vec B_\xi\right)\hat{1} + \frac{2 i}{c} \sum_{k=1}^3 (dt\vec E_\xi \times dt\vec B_\xi)_k\, \hat \sigma_k \nonumber \\
    &= \frac{2dt}{\gamma} \sqrt{\frac{\mu_0}{\epsilon_0}}\left(\frac{\mathcal{E}_\xi}{c}\hat{1} + \sum_k\frac{(\vec{S}_\xi)_k}{c^2}\,\hat{\sigma}_k\right),
\end{align}
proportional to an ensemble-averaged energy density $\mathcal{E}_\xi$ and (Poynting-vector) momentum density $\vec{S}_\xi/c^2$ for the fluctuations, up to a scaling rate $\gamma$; and, (ii) the (frame-independent) complex scalar with real part equal to the fluctuation Lagrangian \cite{dressel_spacetime_2015}, to first order in $dt$,
\begin{align}\label{eq:noise-square}
    (dt\hat{F}_\xi)^2 = (dt \vec E_\xi \cdot dt\vec E_\xi/c^2 - dt \vec B_\xi \cdot dt \vec B_\xi + 2i dt\vec E_\xi \cdot dt \vec B_\xi/c) \hat 1 
    = \frac{dt}{\gamma} (\rho_\xi e^{i\phi_\xi})^2 \hat{1},
\end{align}
with ensemble-averaged real magnitude $\rho_\xi$ and phase $\phi_\xi$, up to the same scaling rate $\gamma$. If $\hat F_\xi$ arises from a single white noise process, then it takes the canonical form \cite{dressel_spacetime_2015},
\begin{align}\label{eq:F-Sigma}
    dt \hat F_\xi &= \hat \Sigma dW / \sqrt{\gamma}, &
    \hat \Sigma &= \rho_\xi \hat f_\xi e^{i \phi_\xi},
\end{align}
where $\hat f_\xi^2 = \hat I$, and $\hat \Sigma$ is a standard-deviation operator satisfying
\begin{align}
    \hat \Sigma^\dagger \hat \Sigma/2 &= \sqrt{\frac{\mu_0}{\epsilon_0}} \left(\frac{\mathcal E_\xi}{c} \hat 1 + \sum_k \frac{(\vec S_\xi)_k}{c^2} \hat \sigma_k\right), & 
    \hat \Sigma^2 &= (\rho_\xi e^{i \phi_\xi})^2\hat{1}.
\end{align}

For qubit measurement Eq.~\eqref{eq:stochastic-me}, informational backaction for the measurement corresponds to a fluctuating electric field. Such a single-component pure electric field fluctuation has covariance proportional to the energy density $\mathcal{E}_\xi$, and since $\hat{B}_\xi = 0$ for a purely electric field fluctuation, the field carries no Poynting momentum $\vec S_\xi \propto \vec E_\xi \times \vec B_\xi = 0$. This lack of Poynting momentum is notable, since it indicates a preferred frame for the fluctuating electromagnetic field in the point charge correspondence. That is, a different inertial observer would see the same fluctuating field differently after a passive Lorentz frame transformation,
\begin{align}
\hat{F}_\xi \mapsto \hat F_\xi' = \,\hat{R}\hat{F}_\xi\hat{R}^{-1} = \hat E_\xi'/c + i \hat B_\xi',
\end{align}
with an apparent magnetic component $\hat{B}_\xi'$ that would yield a non-zero Poynting momentum $\vec S_\xi' \propto \vec E_\xi' \times \vec B_\xi' \neq \vec{0}$. Since an informational measurement process is what breaks the rotational symmetry of a maximally mixed state in the qubit Bloch sphere, it is no surprise that the field fluctuations corresponding to informational measurement in the spacetime generalization have a preferred frame with broken Lorentz symmetry.

Non-informational (unitary) measurement backaction on a qubit corresponds to a fluctuating magnetic field, while more generally, measurement backaction will take the form Eq.~\eqref{eq:F-Sigma} for possibly time-dependent $\phi_\xi$. Interestingly, electric-magnetic duality phase shifts $\hat F_\xi \mapsto \hat F_\xi e^{i\theta}$ \cite{dressel_spacetime_2015,burnsAcousticElectromagneticField2020a,burns2024spacetime} as seen in the qubit measurement of Eq.~\eqref{eq:stochastic-me} preserve the energy-momentum density in Eq.~\eqref{eq:noise-poynting}, but rotate the phase of the complex Lorentz scalar in Eq.~\eqref{eq:noise-square} and thus change the character of the preferred-frame fluctuations. We explore the nontrivial consequences of such a duality phase shift in the next section. Null fluctuations that satisfy $(dt\hat{F}_\xi)^2 = 0$, such as $dt \hat F_\xi = \rho_\xi (\hat \sigma_x + i \hat \sigma_y) dW / \sqrt{\gamma}$, have degenerate phase and no preferred frame. Notably, fluctuations based on electromagnetic waves, such as semi-classical models of vacuum fluctuations \cite{boyerGeneralConnectionRandom1975, goedeckeStochasticElectrodynamicsStochastic1983, huangDiscreteExcitationSpectrum2015, drummondQfunctionsModelsPhysical2020}, are null in character. Since they have no preferred frame, such null fluctuations do not correspond to diffusive qubit measurements of a particular basis as in Eq.~\eqref{eq:stochastic-me}.

Expanding Eq.~\eqref{eq:stochastic-lorentz-transformation} to first order in $dt$ yields a stochastic momentum increment, which similarly decomposes into a mean deterministic part and a zero-mean stochastic part,
\begin{align}\label{eq:stochastic-momentum}
    d \hat p 
    = d \hat p_D + d \hat p_\xi 
\end{align}
with $d\hat{p}_D$ of order $dt$ and $d\hat{p}_\xi$ of order $\sqrt{dt}$ (for details, see the \hyperref[sec:appendix-stochastics]{Appendix}), such that
\begin{align}
        \label{eq:stochastic-dw-dt}
        d \hat p_D &= \frac{\mu}{\hbar} \left\{ dt\hat F_D, \hat p \right\} + \frac{\mu^2}{2\hbar^2} \left\{ dt \hat F_\xi, \left\{ dt \hat F_\xi, \hat p \right\} \right\}, &
        d \hat p_\xi &= \frac{\mu}{\hbar} \{ dt \hat F_\xi, \hat p \}.
\end{align}
The first term in $d \hat p_D$ is the same deterministic term seen in Eq.~\eqref{eq:lorentz-matrix}. The additional double anti-commutator term is a consequence of the variance of the fluctuations influencing the mean dynamics, and has been called an `anti-dephasing' dissipator in the context of a stochastic extension to non-Hermitian quantum theory \cite{martinez-azconaQuantumDynamicsStochastic2025}.

Expanding the normalized velocity increment to linear order in $dt$ similarly yields,
\begin{align}\label{eq:normalized-increment}
    d \hat \beta = d \hat \beta_D + d \hat \beta_\xi,
\end{align}
where the stochastic contribution (derived in the \hyperref[sec:appendix-stochastics]{Appendix}),
\begin{align}\label{eq:normalized-stochastic}
    \begin{aligned}
    d \hat \beta_\xi = \frac{\mu}{\hbar} \bigg(\left\{ dt \hat F_\xi, \hat \beta \right\} - \Tr((dt \hat F_\xi + dt \hat F_\xi^\dagger) \hat \beta) \hat \beta\bigg) 
    \end{aligned}
\end{align}
coincides with the innovation term of qubit stochastic master equations \cite{JacobsIntroContMeas06} like Eq.~\eqref{eq:stochastic-me}. The form of the deterministic normalized increment (derived in the \hyperref[sec:appendix-stochastics]{Appendix}),
\begin{align}\label{eq:normalized-deterministic} 
    d \hat \beta_D &= \frac{\mu}{\hbar} \bigg( \left\{ dt \hat F_D', \hat \beta \right\} - \Tr(((dt \hat F_D') + (dt\hat{F}_D')^\dagger) \hat \beta) \hat \beta\bigg)\nonumber \\
&\qquad    + \frac{\mu^2}{\hbar^2}\bigg((dt \hat F_\xi) \,\hat \beta (dt \hat F_\xi)^{\dagger} - \Tr((dt \hat F_\xi)^\dagger (dt \hat F_\xi) \hat \beta) \hat \beta\bigg),
\end{align}
has two modifications from the deterministic case in Eq.~\eqref{eq:lorentz-normalized-matrix} that arise from the influence of the fluctuation variance. The first term in Eq.~\eqref{eq:normalized-deterministic} has the same form as Eq.~\eqref{eq:lorentz-normalized-matrix}, but includes an effective renormalization of the deterministic generator,
\begin{align}\label{eq:normalized-deterministic-Fd}
    dt \hat F_D' &= dt \hat F_D - \frac{\mu}{\hbar} \Tr( (dt \hat F_\xi + (dt \hat F_\xi)^\dagger) \hat \beta) dt \hat F_\xi.
\end{align}
The second term in Eq.~\eqref{eq:normalized-deterministic} is closely related to the Lindblad ensemble-dephasing term familiar from qubit master equations \cite{Gambetta2006} like Eq.~\eqref{eq:stochastic-me}.

Notably, the quadratic fluctuation corrections to the deterministic increment $dt\hat \beta_D$ contain terms that are nonlinear in $\hat \beta$ due to the renormalization of $\hat \beta$ after each increment. In contrast, the unnormalized momentum increment $dt\, \hat p_D$ remains linear in $\hat p$ since the fundamental group transformation is linear. Since the deterministic mean evolution of a normalized qubit state $\hat \rho$ (analogous to $\hat \beta$) should still be linear in the normalized state $\hat \rho$, as seen in Eq.~\eqref{eq:stochastic-me}, we conclude that qubit measurements must have additional constraints for their corresponding fluctuating fields $dt\hat{F}_\xi$.

The deterministic mean evolution of $\hat{\beta}$ can be nonlinear due to three renormalization terms: the terms $\Tr\left(dt\hat{F}_\xi^\dagger dt\hat{F}_\xi\hat{\beta}\right)\hat{\beta}$ and $\Tr\left((dt\hat{F}'_D + dt\hat{F}'^\dagger_D)\hat{\beta}\right)\hat{\beta}$ in Eq.~\eqref{eq:normalized-deterministic}, and the term $\operatorname{Tr}\left((dt\hat{F}_\xi +  dt\hat{F}_\xi^\dagger\hat{\beta}\right)dt \hat{F}_\xi$ in Eq.~\eqref{eq:normalized-deterministic-Fd}. Additional constraints on the fluctuating fields can make these terms linear in $\hat{\beta}$, as expected for the analogous qubit measurement. Noting that $\Tr \hat{\beta} = 1$, trace terms of the form $\operatorname{Tr}(\hat A \hat \beta)$ will be independent of $\hat \beta$ if $\hat A \propto \hat 1$ and independent of $\hat \beta$. Assuming this sufficient condition for the first two renormalization terms produces the following constraints on the fluctuating fields:
\begin{align}
    \label{eq:poynting-vector-constraint}
    [\hat E_\xi, \hat B_\xi] &= 0 \qquad \leftrightarrow \qquad \vec S_\xi \propto  \vec E_\xi \times \vec B_\xi = 0, \\
    \label{eq:electric-field-constraint}
    \hat E_D' = c\left(\frac{\hat{F}'_D +  \hat{F}'^\dagger_D}{2}\right) &= 0 \qquad \leftrightarrow \qquad dt \vec E_D = \frac{2\mu}{\hbar c} (\vec \beta \cdot dt \vec E_\xi) dt \vec E_\xi.
\end{align}
To compensate for the nonlinearity caused by the third term, we constrain the total renormalized generator in Eq.~\eqref{eq:normalized-deterministic-Fd} to be independent of $\hat \beta$. Since its Hermitian part $\hat{E}'_D = 0$ from Eq.~\eqref{eq:electric-field-constraint}, this remaining constraint on the anti-Hermitian part yields,
\begin{align}
    \label{eq:magnetic-field-constraint}
    \hat B_D' = \frac{\hat{F}'_D - \hat{F}'^\dagger_D}{2i} = dt \hat B_0 \qquad \leftrightarrow \qquad dt \vec B_D = dt \vec B_0 + \frac{2\mu}{\hbar c} (\vec \beta \cdot dt \vec E_\xi) dt \vec B_\xi,
\end{align}
for some $\hat \beta$-independent deterministic magnetic generator $\vec B_0$. 

Under these three conditions, the deterministic mean dynamics of $\hat{\beta}$ in Eq.~\eqref{eq:normalized-deterministic} become linear and precisely reduce to the Lindblad form expected from the qubit master Eq.~\eqref{eq:stochastic-me},
\begin{align}\label{eq:linear-normalized-deterministic} 
    \begin{aligned}
    d \hat \beta_D &= \frac{i \mu dt}{\hbar} \left[ \hat B_0, \hat \beta \right] + \frac{\mu^2}{\hbar^2}\bigg(dt \hat F_\xi \,\hat \beta dt \hat F_\xi^{\dagger} - \frac{1}{2} \left\{dt \hat F_\xi^\dagger dt \hat F_\xi, \hat \beta\right\}\bigg).
    \end{aligned}
\end{align}
Expanding these mean dynamics in 3-vector notation yields,
\begin{align}
    d \vec \beta_D = \frac{ec dt}{p_0} \vec \beta \times \vec B_0 + \frac{e^2c^2}{4 p_0^2} \left((\vec \beta \times dt \vec E_\xi/c) \times dt \vec E_\xi/c + (\vec \beta \times dt \vec B_\xi) \times dt \vec B_\xi \right),
\end{align}
substituting $\mu = e \hbar c / 2 p_0$. The corresponding mean momentum evolution under the same linearity conditions takes the form,
\begin{align}\label{eq:drifting-momentum}
    d \hat p_D = \frac{i \mu dt}{\hbar} \left[\hat B_0, \hat p\right] 
    + \frac{2\mu^2}{\hbar^2 c} \Tr(\hat \beta dt \hat E_\xi) \left\{dt \hat F_{\xi}, \hat p \right\} + \frac{\mu^2}{2\hbar^2}\left\{ dt \hat F_{\xi}, \left\{ dt \hat F_{\xi}, \hat p \right\} \right\}.
\end{align}
or in vector form,
\begin{align}\label{eq:drifting-momentum-vector}
    \begin{aligned}
    d \vec p_D &= 
    e \vec v \times \left( dt \vec B_0 
    + e (\vec p \cdot dt \vec E_\xi) dt \vec B_\xi \right) 
    + \frac{e^2 c}{2 p_0} \left(
        \frac{3}{c^2} \left(\vec v \cdot dt \vec E_\xi\right) dt \vec E_\xi 
        + \left(\vec v \cdot dt \vec B_\xi\right) dt \vec B_\xi + \, dt \vec E_\xi \times dt \vec B_\xi - \vec v \, (dt \vec B_\xi)^2
    \right)\\
    d p_0^D &= 
    \frac{e^2}{p_0 c} \left(
        \vec v \cdot d \vec E_\xi\right)^2 
        + \frac{e^2}{2p_0}\left((dt \vec E_\xi)^2 
        - \vec v \cdot (dt \vec E_\xi \times dt \vec B_\xi)
    \right).
    \end{aligned}
\end{align}
Interestingly, the mean momentum dynamics are reminiscent of runaway self-force solutions: the greater the particle's momentum, the faster it is accelerated. 

Returning to the meaning of the constraints, the first constraint to achieve linearity in Eq.~\eqref{eq:poynting-vector-constraint} states that the Poynting momentum of the zero-mean fluctuations vanishes, which is equivalent to choosing the laboratory frame to be the preferred frame of those fluctuations. The second and third constraints of Eqs.~\eqref{eq:electric-field-constraint} and \eqref{eq:magnetic-field-constraint} together imply that the stochastic electromagnetic fields must have an unusual \emph{velocity-dependent} form, 
\begin{align}\label{eq:lindblad-generator}
    dt \hat F 
    = \frac{1}{c}\left(\frac{2 \mu}{\hbar c} \Tr(dt \hat E_\xi \hat \beta) dt \hat E_\xi + dt \hat E_\xi\right) + i \left(dt \hat B_0 + \frac{2 \mu}{\hbar c} \Tr(dt \hat E_\xi \hat \beta) dt \hat B_\xi + dt \hat B_\xi\right),
\end{align} 
or, equivalently,
\begin{align}
    \begin{aligned}
        dt \vec E &= \frac{2 \mu}{\hbar c} \left(dt \vec E_\xi \cdot \vec \beta\right) dt \vec E_\xi + dt \vec E_\xi \qquad &
        dt \vec B &= dt \vec B_0 + \frac{2\mu}{\hbar c} \left(dt \vec E_\xi \cdot \vec \beta\right) dt \vec B_\xi + dt \vec B_\xi.
     \end{aligned}
\end{align}
As seen in Eq.~\eqref{eq:lorentz-normalized-matrix} for the deterministic case, the fluctuation-independent field $\hat{B}_0$ must be purely magnetic in character for the velocity evolution $d\hat{\beta}_D$ to be linear in $\hat{\beta}$, meaning that the velocity must not be boosted on average. However, only fluctuations with a velocity-dependent mean of the specific form in Eq.~\eqref{eq:lindblad-generator} can preserve that linearity by compensating for the mean boost induced by the fluctuations. The need for this compensating term can thus be understood as a stabilization condition to prevent boosts of the mean reference frame of the charge (i.e., the mean $\hat{\beta}$). Moreover, the form of Eq.~\eqref{eq:lindblad-generator} implies that observing such a fluctuating electromagnetic field would constitute a noisy measurement of a particular velocity component, in complete analogy to how observing the noisy record $r$ for a continuous qubit measurement as in Eq.~\eqref{eq:stochastic-me} constitutes a noisy measurement of a particular qubit observable (which we will carefully consider in the next section). It is curious and notable that demanding linearity of the mean evolution of $\hat{\beta}$ is a sufficient condition to establish such a close correspondence to qubit measurement.

\begin{table}[H]
    \def\arraystretch{1.5}
      \centering\small
    
    \begin{tabular}{l|l}
        \multicolumn{1}{c}{\textbf{Point Charge}} & \multicolumn{1}{c}{\textbf{Qubit}} \\ \hline \hline \multicolumn{2}{c}{}\vspace{-1em}\\
        \multicolumn{2}{c}{\textbf{Stochastic Spacetime (Unnormalized) Dynamics}} \\ \hline
        \multicolumn{2}{c}{Stochastic Generators} \\
        $\hat F = \hat F_D + \hat F_\xi$ 
        & 
        $\hat H_\text{eff} = \hat H_D + \hat H_\xi$
        \\
        \hline
        \multicolumn{2}{c}{Stochastic Lorentz Transformation} \\
        $\begin{aligned}
            \hat p(t + dt) &= e^{\mu dt \hat F/\hbar} \hat p(t) e^{\mu dt \hat F^\dagger/\hbar} \approx \hat p(t) + d \hat p_D + d \hat p_\xi
        \end{aligned}$ 
        & 
        $\begin{aligned}\hat s(t + dt) &= e^{-i dt \hat H_\text{eff}/\hbar} \hat s(t) e^{i dt \hat H_\text{eff}^\dagger/\hbar} \approx \hat s(t) + d \hat s_D + d \hat s_\xi\end{aligned}$ 
        \\ \hline
        \multicolumn{2}{c}{Innovation} \\
        $d \hat p_\xi = \mu \{dt \hat F_\xi, \,\hat p\}/\hbar$
        &
        $d \hat s_\xi = \{-i dt \hat H_\xi, \, \hat s \}/\hbar$ 
        \\ \hline
        \multicolumn{2}{c}{Mean Dynamics} \\
        $d \hat p_D = \mu \, dt \{ \hat F_D, \hat p \}/\hbar + \mu^2 \{ dt \hat F_\xi, \{ dt \hat F_\xi, \hat p \} \} / 2\hbar^2$
        &
        $d \hat s_D = -i dt [\hat H_D, \hat s]/\hbar + \{ dt \hat H_\xi, \{ dt \hat H_\xi, \hat s \} \}/2\hbar^2$ 
        \\ \multicolumn{2}{c}{}\\
        \multicolumn{2}{c}{\textbf{Stochastic Bloch Ball (Normalized) Dynamics}} \\ \hline
        \multicolumn{2}{c}{Normalized Stochastic Evolution} \\
        $\hat \beta(t + dt) = \hat p(t + dt)/\Tr(\hat p(t + dt)) \approx \hat \beta(t) + d \hat \beta_D + d \hat \beta_\xi$ 
        & 
        $\hat \rho(t + dt) = \hat s(t + dt)/\Tr(\hat s(t + dt)) = \hat \rho(t) + d \hat \rho_D + d \hat \rho_\xi$ 
        \\
         \hline
         \multicolumn{2}{c}{Innovation} \\
        $d \hat \beta_\xi = \mu (\{dt \hat F_\xi, \hat \beta \} - \Tr(\{dt \hat F_\xi, \hat \beta \}) \hat \beta)/\hbar$ & $d \hat \rho_\xi = \{-i dt \hat H_\xi, \hat \rho \} - \Tr(\{-i dt \hat H_\xi, \hat \rho \}) \hat \rho$ \\
        \hline
         \multicolumn{2}{c}{Mean Dynamics} \\
         $\begin{aligned}
         d \hat \beta_D &= \mu\bigg( \{ dt \hat F_D', \hat \beta \} - \Tr(\{ dt \hat F_D', \hat \beta \}) \hat \beta\bigg)/\hbar \\
         &+ \mu^2\bigg(dt \hat F_\xi \,\hat \beta dt \hat F_\xi^{\dagger} - \Tr(dt \hat F_\xi \hat \beta dt \hat F_\xi^\dagger) \hat \beta\bigg)/\hbar^2
         \end{aligned}$ & $\begin{aligned}
         d \hat \rho_D &= \bigg( -i[dt \hat H_D', \hat \rho ] - \Tr(-i[dt \hat H_D', \hat \rho ]) \hat \rho\bigg)/\hbar \\
         &+ \bigg( dt \hat H_\xi \hat \rho dt \hat H_\xi^\dagger - \Tr(dt \hat H_\xi \hat \rho dt \hat H_\xi^\dagger)\hat \rho \bigg)/\hbar^2
         \end{aligned}$ \\
         \hline
         \multicolumn{2}{c}{Effective Generator} \\
         $dt \hat F_D' = dt \hat F_D - \Tr(\mu \{dt \hat F_\xi, \hat \beta \}/\hbar) dt \hat F_\xi$ & $dt \hat H_D' = dt \hat H_D - \Tr(\{ -i d \hat H_\xi, \hat \rho \}/\hbar) d \hat H_\xi$ 
    \end{tabular}
    
    \caption{Summary of the stochastic correspondence between a relativistic point charge and a qubit, where $dt \hat F_\xi$ and $dt \hat H_\xi$ are order $\sqrt{dt}$. Unnormalized quantities exist in a 4-dimensional spacetime, while normalized quantities exist in a 3-dimensional Bloch ball. The commutator and anticommutator are given by $[\hat A, \hat B] := \hat A \hat B - \hat B \hat A^\dagger$ and $\{ \hat A, \hat B\} = \hat A \hat B + \hat B \hat A^\dagger$, respectively.
    }\label{tab:stochastic-correspondence}
\end{table}

\section{Delayed Choice Lorentz Transformations} \label{sec:delay}

For concreteness, we now revisit the $\theta$-dependent continuous Gaussian measurement of a qubit with the measurement operator that we previously introduced in Eq.~\eqref{eq:gaussian-theta},
\begin{align}\label{eq:lorentz-theta}
    \hat{R}_{r,\theta} &= \exp(\Gamma \Delta t\, r e^{-i\theta}\,\hat{\sigma}_z).
\end{align} 
We explain in more detail how the parameter $\theta$ relates to a delayed choice being made in an experimental implementation, then discuss the implications of such a parameter for the analogous point-particle correspondence.

The measurement operator $\hat{R}_{r,\theta}$ is a good description for the typical readout process of a superconducting qubit, such as a transmon or fluxonium, using circuit quantum electrodynamics (cQED) \cite{Koch2007}. In such a superconducting circuit setting, the qubit consists of the lowest two energy levels of an anharmonic oscillator (with energy gap $E_1 - E_0 = \hbar\omega_q$) and is coupled to a harmonic microwave readout resonator of bare frequency $\omega_r$ with amplitude $\hat{a}$ (satisfying $[\hat{a},\hat{a}^\dagger] = \hat{1}$) that is kept strongly detuned from the qubit. Such a dispersive coupling causes the frequency of the resonator to depend on the qubit state, with an effective interaction Hamiltonian $\hat H_\text{int} = -\hbar \chi \hat a^\dagger \hat a \hat \sigma_z$, where $\pm\chi$ is the dispersive qubit-dependent frequency shift of the resonator. The readout resonator then couples to a transmission line with amplitude decay rate $\kappa/2$, allowing an input microwave tone to reflect off the resonator then be amplified and recorded with a homodyne measurement to produce a stochastic time series $(r_0,r_1,r_2,\ldots)$ of digitized readout signals $r_k$ separated by an integration time steps $\Delta t$ corresponding to the inverse bandwidth of the homodyne detector. Each readout $r_k$ in the time series corresponds to a measurement operator $\hat{R}_{r_k,\theta_k}$, where $\theta_k$ corresponds to the quadrature angle of the output microwave tone that was probed by the homodyne measurement to produce the result $r_k$.

In more detail, pumping the transmission line with a microwave tone $\varepsilon e^{-i\omega_r t}$ at the bare resonance frequency $\omega_r$ adiabatically generates two distinct coherent steady states in the resonator with amplitudes $\alpha_\pm = -i(2\varepsilon/\kappa)/(1 \mp i(2\chi/\kappa))$, each with the same mean photon number $\bar{n} = |\alpha_\pm|^2 = |2\varepsilon/\kappa|^2/(1 + (2\chi/\kappa)^2)$ but differing phase shifts $\text{arg}(\alpha_\pm) = -\pi/2\pm\arctan(2\chi/\kappa)$. The different $\theta$-dependent resonator quadratures $\hat{I}_\theta = (\hat{a}e^{-i\theta} + \hat{a}^\dagger e^{i\theta})/\sqrt{2}$ will thus have varying information about the qubit state. One extreme case is $\hat{I}_0 = (\hat{a}+\hat{a}^\dagger)/\sqrt{2}$, which will have two distinct steady-state means $\text{Re}(\alpha_\pm) = \pm\bar{n}(\chi/\varepsilon)$ and thus reveals maximum information about the qubit state. The other extreme case is $\hat{I}_{\pi/2} = -i(\hat{a} -\hat{a}^\dagger)/\sqrt{2}$, which only has a single steady-state mean $\text{Im}(\alpha_\pm) = -\bar{n}(\kappa/2\varepsilon)$ and thus reveals no information about the qubit state. Since the resonator state leaks into the transmission line and is later measured, we thus expect distinct types of measurement backaction on the qubit that depend on the measured quadrature angle $\theta$. 

The qubit and resonator cannot know what quadrature will be measured in the transmission line at a later time after it propagates to the homodyne detector. This no-signaling constraint forces the ensemble-average dynamics to be $\theta$-independent. To see this formally, note that over a duration $\Delta t$ each coherent state of the resonator leaks into the transmission line with probability $\kappa\Delta t$, yielding an entangled steady state \cite{Korotkov2016} (where we order the states as $\ket{\rm qubit}\ket{\rm resonator}\ket{\rm{transmission~line}}$ and work in a rotating frame that cancels the additional state-dependent phase evolution from the drive),
\begin{align}\label{eq:entangled-state}
    \ket{\Psi} &= c_0\ket{0}\ket{\alpha_+}\ket{\sqrt{\kappa \Delta t}\,\alpha_+} + c_1\ket{1}\ket{\alpha_-}\ket{\sqrt{\kappa \Delta t}\,\alpha_-}.
\end{align}
The reduced qubit-resonator state after tracing out the leaked field in the transmission line,
\begin{align}
    \hat{\rho}_{\rm qr} &= |c_0|^2\ket{0}\!\bra{0}\otimes\ket{\alpha_+}\bra{\alpha_+} + |c_1|^2\ket{1}\!\bra{1}\otimes\ket{\alpha_-}\!\bra{\alpha_-} + e^{-\Gamma \Delta t}(c_0c_1^* e^{i\omega_S\Delta t}\ket{0}\!\bra{1}\otimes\ket{\alpha_+}\!\bra{\alpha_-} + \text{h.c.}),
\end{align}
thus exhibits ensemble-averaged dephasing and phase precession,
\begin{align}
    e^{-\Gamma\Delta t +i\omega_S\Delta t} &= \braket{\sqrt{\kappa \Delta t}\,\alpha_-}{\sqrt{\kappa\Delta t}\,\alpha_+} = e^{-\kappa \Delta t|\alpha_- - \alpha_+|^2/2 + i\kappa \Delta t\,\text{Im}(\alpha_-^*\alpha_+)},
\end{align}
at the \emph{measurement-dephasing rate} and \emph{ac-Stark-shift frequency} (up to the omitted drive-induced phase difference) \cite{Gambetta2006,Gambetta2008,Korotkov2016},
\begin{align}\label{eq:gamma}
    \Gamma &= \kappa\frac{|\alpha_- - \alpha_+|^2}{2} = \frac{2}{\kappa}\frac{(2\chi)^2\bar{n}}{1 + (2\chi/\kappa)^2}, &
    \omega_S &= \kappa\,\text{Im}(\alpha_-^*\alpha_+) = \frac{4\chi\bar{n}}{1 + (2\chi/\kappa)^2}.
\end{align}
This reduced state evolution that traces out the transmission line is equivalent to ensemble-averaging over all results measured later on the transmission line. Thus, any choice of quadrature angle $\theta$ that is later measured will produce precisely the same ensemble-averaged evolution for the qubit-resonator state.

\begin{figure}[H]
    \centering
{\includegraphics[width=0.9\linewidth]{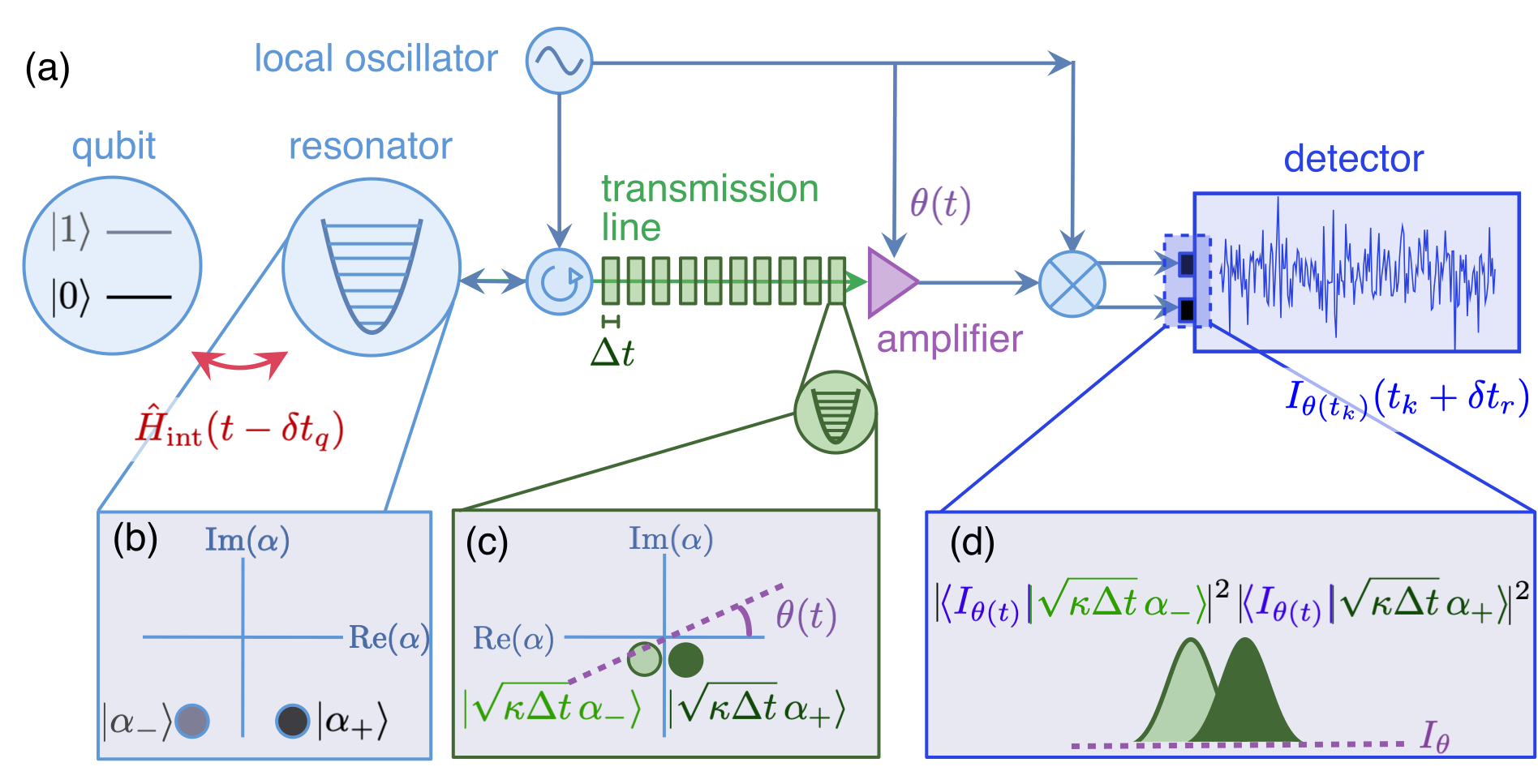}}
    \caption{A superconducting qubit is dispersively coupled to a resonator via $\hat H_\mathrm{int}(t - \delta t_q)$ (a). This causes the entangled qubit-resonator state to reach a steady-state superposition of $\ket{0}\ket{\alpha_+}$ and $\ket{1}\ket{\alpha_-}$, sketched in (b) as the Husimi-Q function $Q(\alpha)$ of the conditioned resonator states. The resonator is pumped in reflection via the local oscillator, which entangles the qubit-resonator states with traveling transmission line states $\ket{\sqrt{\kappa \Delta t} \, \alpha_\pm}$ discretized by the detector integration time $\Delta t$ (c). The traveling modes are amplified along a quadrature at angle $\theta(t)$ that can be freely and continuously varied throughout the experiment, which determines whether the qubit undergoes unitary ($\theta = \pi/2$) backaction, non-unitary hyperbolic ($\theta = 0$) backaction, or a combination of both.  The mixer multiplies the leaked, amplified resonator field with the LO and low-pass filters the output,
     yielding a slowly varying homodyne signal $I_{\theta(t)}(t + \delta t_r)$ randomly distributed as a mixture of two real Gaussians due to postselection of the transmission line modes $\ket{\sqrt{\kappa\Delta t}\, \alpha_\pm}$ by the tilted quadrature eigenstate $\bra{I_{\theta(t)}}$ (d). This observation at time $t + \delta t_r$ collapses the entangled state, retroactively determining the reduced qubit state at the prior original interaction time $t-\delta t_q$ as a result of the choice of amplification quadrature at time $t$.}
\label{fig:circuit}
\end{figure}

For the actual measurement procedure, shown in Fig.~\ref{fig:circuit}, the leaked field propagates down the transmission line unchanged for a time delay $\delta t_q$, then at time $t$ passes through a phase-sensitive amplifier that amplifies the quadrature defined by angle $\theta(t)$ and deamplifies the complementary quadrature $\theta(t) + \pi/2$. Equivalently, this can be understood as squeezing the traveling coherent state $\ket{\sqrt{\kappa \Delta t} \alpha_\pm}$ along an axis $\theta(t)$ in the complex plane of $\alpha_\pm$. Importantly, the quadrature angle $\theta(t)$ is determined by the relative phase between the amplifier reference pump and the traveling field to be amplified, so can be made long after the traveling field already interacted with the qubit and resonator. The amplified field is then downconverted by mixing with the reference pump and digitized to produce the stochastic homodyne readout $I_{\theta(t)}(t+\delta t_r)$ after another time delay $\delta t_r$. The total time delay from the time of interaction to the completion of the readout procedure is $\delta t_q + \delta t_r$, so the measured readout will correspond to (and affect) the \emph{past} state of the qubit and resonator. After a sequence of homodyne measurements $I_{\theta(t_k)}(t_k + \delta t_r)$ have been recorded, the pump is turned off to adabiatically return the readout resonator to the vacuum state, which disentangles the resonator from the qubit to leave  a final conditioned qubit state that is entirely determined by the recorded history of measurements.

The collected definite signal $I_{\theta(t)}(t+\delta t_r)$ is proportional to the eigenvalue of the corresponding transmission line quadrature $\hat{I}_{\theta(t)} = (\hat{c}e^{-i\theta(t)}+\hat{c}^\dagger e^{i\theta(t)})/\sqrt{2}$, where $\hat{c}$ is the amplitude of the transmission line mode containing the leaked (and amplified) field. Thus, measuring the result $I_{\theta(t)}(t+\delta t_r)$ at time $t+\delta t_r$ for the quadrature angle $\theta$ chosen at time $t$ collapses the entangled state generated by the interaction at time $t-\delta t_q$ in Eq.~\eqref{eq:entangled-state} to leave a conditioned (unnormalized) state of the qubit and resonator:
\begin{align}\label{eq:entangled-state-collapse}
    \ket{\psi_{I_{\theta(t)}(t+\delta t_r)}(t-\delta t_q)} &= c_0\ket{0}\ket{\alpha_+}\braket{I_{\theta(t)}(t+\delta t_r)}{\sqrt{\kappa \Delta t}\,\alpha_+} + c_1\ket{1}\ket{\alpha_-}\braket{I_{\theta(t)}(t+\delta t_r)}{\sqrt{\kappa \Delta t}\,\alpha_-} \nonumber \\
    &= \left(\hat{M}_{I_{\theta(t)}(t+\delta t_r)}\otimes\hat{1}\right)\ket{\psi(t-\delta t_q)}, \\
    \hat{M}_{I_{\theta(t)}(t+\delta t_r)} &= \braket{I_{\theta(t)}(t+\delta t_r)}{\sqrt{\kappa \Delta t}\,\alpha_+}\,\ket{0}\!\bra{0} + \braket{I_{\theta(t)}(t+\delta t_r)}{\sqrt{\kappa \Delta t}\,\alpha_-}\,\ket{1}\!\bra{1},
\end{align}
where $\ket{\psi(t-\delta t_q)} = c_0\ket{0}\ket{\alpha_+} + c_1\ket{1}\ket{\alpha_-}$ is the joint state of just the qubit and resonator at the time of interaction $t-\delta t_q$, assumed to be adiabatically prepared from the product state $(c_0\ket{0}+c_1\ket{1})\ket{0}$ of a qubit state and the vacuum state of the resonator. The effective measurement operator $\hat{M}_{I_{\theta(t)}(t+\delta t_r)}$ notably affects only the qubit degree of freedom and is diagonal in the $\hat{\sigma}_z$ basis, which is why the intermediate entanglement with the resonator steady states does not impact the final reduced qubit dynamics after adiabatic disentanglement with the resonator at the end of the measurement sequence. This enables the standard Bayesian updates of the form in Eqs.~\ref{eq:M_r0},~\ref{eq:post-measurement-state}, in which efficient quantum measurement appears to maintain purity of a pure qubit state. In fact, the qubit-resonator entanglement causes the reduced qubit state to be mixed during the intervening measurement dynamics. However, the adiabatic assumption forces states of the form $\ket 0 \ket{\alpha_+}$ and $\ket 1 \ket{\alpha_-}$, which can be understood as the true logical states to which Eqs.~\ref{eq:M_r0},~\ref{eq:post-measurement-state} refer, while adibatic disentanglement prior to tomography ensures that this subtlety around the definition of the logical states is not important to experimental prediction of the qubit state evolution.  

The effective measurement operator in Eq.~\ref{eq:entangled-state-collapse} is determined by the collapse amplitudes from the distinct leaked steady states in the transmission line to the observed homodyne signal $I_{\theta(t)}(t+\delta t_r)$. The probability density for the observed signal $I_{\theta(t)}(t+\delta t_r)$ is correspondingly determined by the Born rule, as anticipated in Eq.~\eqref{eq:povm},
\begin{align}
    p(I_{\theta(t)}(t + \delta t_r)) &= \bra{\psi(t-\delta t_q)}\left[\hat{M}^\dagger_{I_{\theta(t)}(t+\delta t_r)}\hat{M}_{I_{\theta(t)}(t+\delta t_r)}\otimes\hat{1}\right]\ket{\psi(t-\delta t_q)}.
\end{align}
Evaluating the collapse amplitudes in $\hat{M}_{I_{\theta(t)}(t+\delta t_r)}$ yields (suppressing the time arguments for brevity),
\begin{align}
    \braket{I_{\theta}}{\sqrt{\kappa \Delta t}\,\alpha_\pm} &= \frac{\exp\left[-\frac{1}{2}\left(I_\theta- \sqrt{2\kappa \Delta t}\text{Re}(\alpha_\pm e^{-i\theta})\right)^2 + iI_\theta \sqrt{2\kappa\Delta t} \text{Im}(\alpha_\pm e^{-i\theta}) - i\kappa\Delta t\text{Re}(\alpha_\pm e^{-i\theta})\text{Im}(\alpha_\pm e^{-i\theta})\right]}{\pi^{1/4}}. 
\end{align}
After recalling the steady state amplitudes $\alpha_\pm = -i\sqrt{\bar{n}}\,\exp(\pm i \arctan(2\chi/\kappa))$ with $\bar{n} = (2\varepsilon/\kappa)^2/(1 + (2\chi/\kappa)^2)$ and identifying the suitably shifted and rescaled result coordinate
\begin{align}
    r &\equiv \frac{I_\theta}{\sqrt{\Gamma\Delta t}} + \frac{\kappa}{2\chi}\sin\theta 
\end{align}
that involves the ensemble-average dephasing rate $\Gamma$ in Eq.~\eqref{eq:gamma}, these collapse amplitudes produce a corresponding probability distribution for $r$
\begin{align}\label{eq:r-prob}
    p(I_\theta)dI_\theta &= p(r)dr = \sqrt{\frac{\Gamma \Delta t}{\pi}}\bra{\psi}\left[\exp\left(-\Gamma\Delta t(r - \cos\theta\,\hat{\sigma}_z)^2\right)\otimes\hat{1}\right]\ket{\psi}\,dr = \bra{\psi}\left[\hat{M}_r^\dagger\hat{M}_r\otimes\hat{1}\right]\ket{\psi}\,dr
\end{align}
that is a mixture of state-dependent Gaussian distributions for the result $r$, each centered at distinct means $\pm\cos\theta$ but with with equal variance $1/2\Gamma\Delta t$, as anticipated in Eqs.~\eqref{eq:gaussian-theta}. After reparameterization with the result coordinate $r$, the corresponding measurement operator takes the form,
\begin{align}
    \hat{M}_r &\equiv (\Gamma \Delta t)^{1/4}\left[\braket{I_\theta}{\sqrt{\kappa\Delta t}\,\alpha_+} \ket{0}\!\bra{0} + \braket{I_\theta}{\sqrt{\kappa\Delta t}\,\alpha_-} \ket{1}\!\bra{1}\right] = \sqrt{\bar{p}(r,\theta)}\,e^{i\omega_S\hat{\sigma}_z/2}\,\hat{R}_{r,\theta},
\end{align}
including a state-independent factor as expected in Eqs.~\eqref{eq:gaussian-theta} that cancels during renormalization,
\begin{align}
    \bar{p}(r,\theta) &= \sqrt{\frac{\Gamma\Delta t}{\pi}}e^{-\Gamma \Delta t\,(r^2 + \cos^2\theta) -2i\Delta t\,\bar{\omega}(r,\theta)}, &
    \bar{\omega}(r,\theta) &= \omega_S\,r\cos\theta - \frac{\Gamma}{4}\left(1 + \left(\frac{\kappa}{2\chi}\right)^2\right)\sin(2\theta),
\end{align}
a unitary deterministic frequency shift $\omega_S$ from the ac-Stark effect in Eq.~\eqref{eq:gamma} that can be compensated by shifting the frequency of the rotating frame of the qubit, and the $r$-dependent stochastic measurement backaction $\hat{R}_{r,\theta}\in\text{SL}(2,\mathbb{C})$ anticipated in Eqs.~\eqref{eq:gaussian-theta} and Eq.~\eqref{eq:lorentz-theta}.

Therefore, after choosing the appropriate rotating frame for the qubit, rescaling the observed homodyne readout $I_\theta$ to $r$, and partially renormalizing the state to cancel state-independent prefactors, the effective measurement backaction induced by the standard readout procedure for superconducting qubits for each time interval $\Delta t$ in a sequence indeed has the form $\hat{R}_{r(t+\delta t_r),\theta(t)}\in\text{SL}(2,\mathbb{C})$ analogous to a Lorentz transformation of a point charge, as claimed. Moreover, for small $\Delta t$ the distribution for the readout result $r$ in Eq.~\eqref{eq:r-prob} approximates a single Gaussian distribution centered at the mean value $\langle r(t+\delta t_r)\rangle = \cos[\theta(t)]\,\text{Tr}[\hat{\sigma}_z\,\hat{\rho}(t-\delta t_q)]$ determined by the expectation value of $\hat{\sigma}_z$ in the reduced qubit state $\hat{\rho}(t-\delta t_q)$ and the choice of homodyne angle $\theta(t)$, with variance $1/2\Gamma\Delta t$. A formal time-continuous limit $\Delta t \to dt$ can then approximate the observed readout as a moving-mean white-noise process,
\begin{align}
    \Delta t\,r(t+\delta t_r) &\to dt\,r(t+\delta t_r) = dt\,\cos[\theta(t)]\,\text{Tr}[\hat{\sigma}_z\hat{\rho}(t-\delta t_q)] + dW(t + \delta t_r)/\sqrt{2\Gamma},
\end{align}
where the zero-mean stochastic Wiener increment $dW$ satisfies the It\^o rule $dW^2=dt$. Using this white-noise process limit in $\hat{R}_{r(t+\delta t_r),\theta(t)}$ for the reduced qubit state update and linearizing to construct a forward difference It\^o derivative then produces the anticipated stochastic differential equation in Eq.~\eqref{eq:stochastic-me}.

The readout result $r(t+\delta t_r)$ depends on the (normalized) qubit state $\hat{\rho}(t-\delta t_q)$ and choice of homodyne angle $\theta(t)$ at earlier times, which is unsurprising. What is more surprising is that this more complete derivation shows that the resulting measurement backaction $\hat{R}_{r(t+\delta t_r),\theta(t)}$ still affects the qubit state $\hat{\rho}(t-\delta t_q)$ at the \emph{earlier} time $t-\delta t_q$. Specifically, the type of conditioned evolution in $\hat{R}_{r(t+\delta t_r),\theta(t)}$ that affects the qubit state at time $t-\delta t_q$ depends on the choice of homodyne angle $\theta(t)$ made at the later time $t$, while the amount of backaction depends on the result $r(t+\delta t_r)$ obtained at the even-later time $t+\delta t_r$. The future choice of $\theta(t)$ will determine whether the backaction is unitary or non-unitary (hyperbolic), yielding orthogonal motions on the Bloch sphere that are readily distinguishable. This delayed choice nature of the dynamics can be (and has been) confirmed by subsequent measurements made on the qubit after observing specific records $r(t+\delta t_r)$ \cite{Murch2013,Weber2014,koolstra2022monitoring}. The detailed state-tracking statistics observed in these experiments are completely consistent with the retrocausal character of the effective backaction derived here. As with all delayed choice effects in quantum mechanics, however, this apparently retrocausal effect does have a more causal interpretation where the specific qubit dynamics are left indefinite as part of a joint entangled state with intrinsic uncertainty like Eq.~\eqref{eq:entangled-state} until the final collapse from observation like in Eq.~\eqref{eq:entangled-state-collapse}. Nevertheless, once the specific record $r(t+\delta t_r)$ is known, the entire past history of the qubit dynamics effectively collapses to be consistent with that future record, making the retrocausal backaction $\hat{R}_{r(t+\delta t_r),\theta(t)}$ an accurate description of the effective state dynamics.

\begin{figure}[H]
    \centering
    \includegraphics[width=\linewidth]{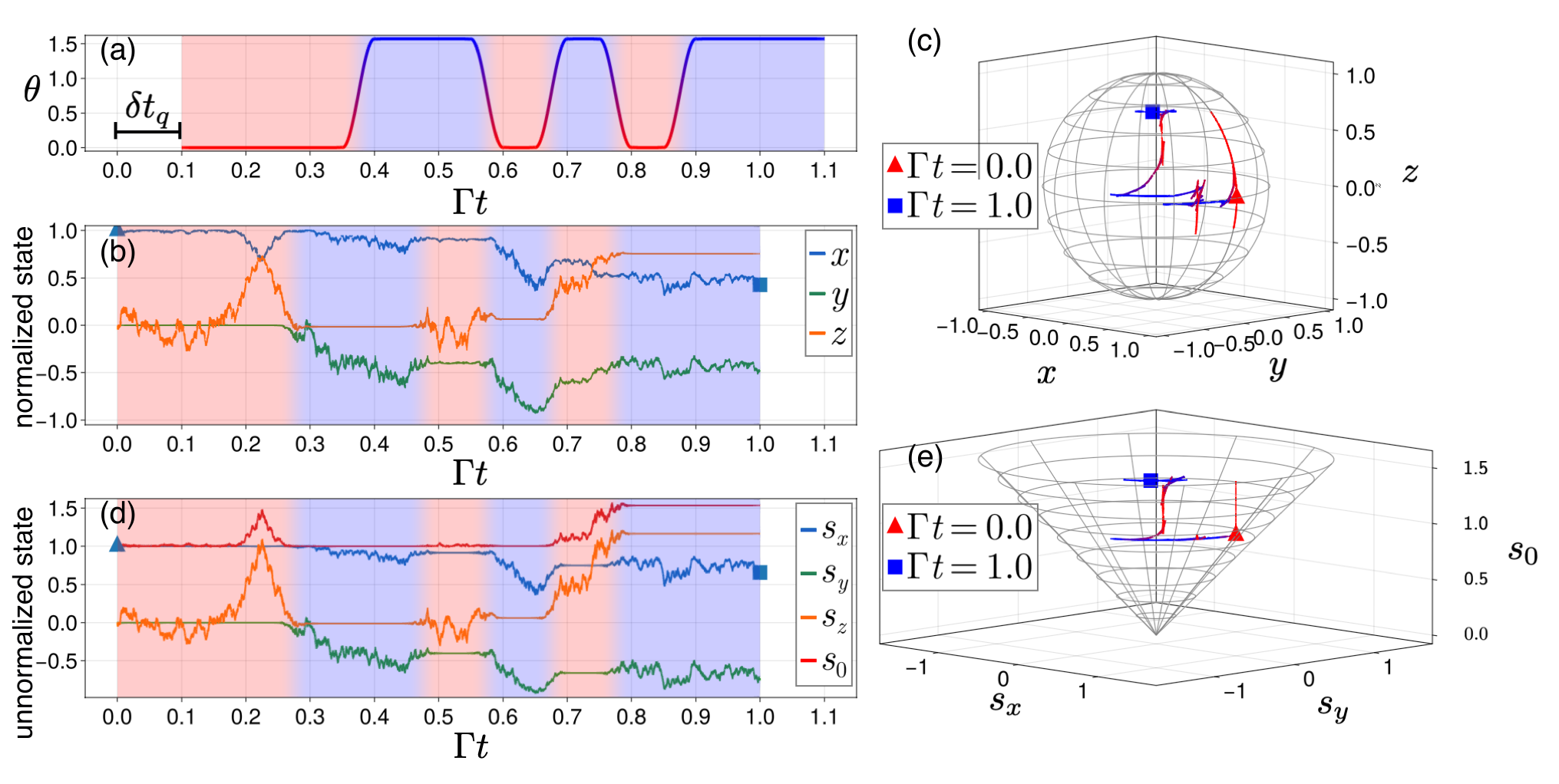}
    \caption{Quantum trajectory simulation depicting the normalized $\hat \rho(t)$ and unnormalized $\hat s(t)$ state evolution of a continuously monitored qubit under dynamical delayed choice of the homodyne (squeezing) angle $\theta(t)$. Equivalently, one may interpret these dynamics as the (normalized) velocity $\hat \beta$ and (unnormalized) momentum $\hat p$ evolution of a point charge in a velocity-dependent stochastic electromagnetic field with duality phase $\theta(t)$, in the form given by Eq.~\eqref{eq:advanced-stochastic-field}.  (a) The squeezing angle is varied smoothly between informational ($\theta = 0$, red) and non-informational ($\theta=\pi/2$, blue) measurement backaction. (b) The time delay $\Gamma \delta t_q = 0.1$ between qubit and amplifier leads the choice of $\theta$ at times $t + \delta t_q$ to retroactively determine the inferred qubit dynamics at times $t$, as witnessed in the Bloch coordinates of the normalized state $\hat \rho$. (c) On the Bloch sphere, informational measurement (red) generates hyperbolic backaction between measurement eigenstates $\ket 0$ and $\ket 1$; non-informational measurement (blue) generates rotational backaction around $\hat \sigma_z$; combined informational and non-informational measurement (purple) leads to a combination of both. The initial and final states are marked by a triangle and square, respectively. (d) The unnormalized state $\hat s$ shows backaction in $s_z$ and $s_0$ for $\theta = 0$ and in $s_x$ and $s_y$ for $\theta = \pi/2$. The absence of stochasticity in the $x,y$ components for $\theta = 0$ (red background) highlights the purely hyperbolic nature of informational backaction, which is otherwise obscured in the timeseries data of $\hat \rho$ by renormalization. (e) The unnormalized state lives on a four-dimensional lightcone, represented here as the $s_x$-$s_y$-$s_0$ slice. Noninformational backaction is visible as rotations in $s_x$-$s_y$, while informational backaction manifests as vertical motion of $s_0$ ($s_z$ is not visible in this slice).}
    \label{fig:stochastics}
\end{figure}

To underscore how unusual this behavior of a monitored qubit state is, despite the formal equivalence to a stochastic Lorentz transformation, consider what would have to happen to achieve equivalent dynamics with a classical point charge according to the correspondence developed in the previous sections. The four-momentum $\hat{p}(t)$ of the charge at a time $t$ would have to experience an external stochastic electromagnetic force,
\begin{subequations}\label{eq:advanced-stochastic-field}
\begin{align}
    \hat p(t+dt) &= e^{\mu dt \hat F(t)/\hbar} \hat p(t) e^{\mu dt \hat F^\dagger(t)/\hbar}, \\
    dt \hat F(t) &= dt\left(\hat{E}(t)/c + i\hat{B}(t)\right) = \frac{E_0}{c} \left(\cos[\theta(t+\delta t_q)] \beta_z(t)\,dt  +  dW(t+\delta t_q+\delta t_r)/\sqrt{2\Gamma}\,\right)  e^{-i \theta(t + \delta t_q)} \hat \sigma_z,
\end{align}
\end{subequations}
with $E_0\mu/\hbar c = \Gamma$ such that the field explicitly depends on $\beta_z(t) = v_z(t)/c = p_z(t)/p_0(t)$, the $z$-component of the particle velocity ratio at time $t$. Such a velocity-dependent external electric field is a form of instantaneous feedback that is difficult to arrange for a classical charge. Moreover, the decomposition of the field into electric and magnetic parts would have to be determined by an experimental phase parameter $\theta(t + \delta t_q)$ that is selected a delay $\delta t_q$ in the future after the field has already interacted with the particle at time $t$, exhibited in Fig.~\ref{fig:stochastics}. Such a delayed choice effect is equivalent to preserving the dual-symmetry (i.e., electric-magnetic egalitarianism) of the electromagnetic field in vacuum long after the interaction with the charge \cite{dressel_spacetime_2015,burnsAcousticElectromagneticField2020a}. The choice $\theta = 0$ would produce a purely electric force in the past, while the choice $\theta = \pi/2$ would produce a purely magnetic force in the past. The fluctuations in the field $dW(t+\delta t_q+\delta t_r)$ would have to be determined even later in the future after an additional delay $\delta t_r$. This velocity dependence and temporal nonlocality that would be needed in the corresponding charge dynamics to mimic the behavior of a monitored qubit cannot be replicated by external electromagnetic fields that obey causal propagation.

\section{Conclusion} \label{sec:conclusion}

In this paper, we generalized the traditional Bloch ball representation of a normalized qubit state as a three-dimensional spin vector by lifting the normalization constraint and extending it to a proper four-vector that behaves analogously to the four-momentum of a point charge in spacetime. Crucially, this spacetime representation is faithful to the enlarged dynamical group $\text{SL}(2,\mathbb{C})$ of a monitored qubit, so can describe aspects of monitored qubit dynamics that are suppressed by the renormalized Bloch vector representation. Indeed, we showed that the Bloch three-vector corresponds in this spacetime representation to the velocity of the corresponding charge in a particular frame, giving an interesting perspective on the apparent nonlinearity that appears in the renormalized monitored qubit dynamics. We thus expect this spacetime correspondence to be a useful visualization tool for monitored qubit dynamics that can provide intuition complementary to the Bloch ball picture. 

A notable feature of the spacetime representation is that deterministic non-Hermitian dynamics are equivalent to the familiar Lorentz force on a charge generated by an electromagnetic field. This equivalence implies that the charge dynamics induced by stochastic electromagnetic fields should be able to emulate the stochastic non-Hermitian dynamics of a monitored qubit, as well as the deterministic non-Hermitian dynamics of a dissipative qubit. However, we showed that establishing the detailed correspondence to monitored qubit dynamics places unusual constraints on the fluctuating force fields: First, the stochastic electromagnetic field must depend on the velocity of the charge it is acting on, making it a fluctuating feedback-control field. Second, whether the electromagnetic field is electric or magnetic in character depends on a parameter that an experimenter can control in the future, making the type of Lorentz transformation felt by the charge a dynamical form of delayed choice. These unusual dynamical features mandated by our classical spacetime charge correspondence help clarify some of the inherent strangeness of monitored qubit evolution. 

The subtle parameter dependence on the delayed future choice of an experimenter particularly highlights the inherent conceptual tension between causality and the entangled-state collapse induced by quantum measurement. We emphasize that such tension cannot be avoided in any classical hidden variable model of the measurement process, including the charge correspondence we develop here. The apparent retrocausality in our charge model thus exemplifies Carmichael's warning in Ref.~\cite{carmichaelOpenSystemsApproach1993} regarding the nature of monitored quantum trajectories: ``If we adopt this [trajectory] viewpoint\ldots we must still accept, or somehow circumvent, a manifest nonlocality in time\ldots [The fluctuations] came from the randomness of photoelectron emissions at the detector, communicated backwards in time to the source by the collapse of the wavefunction.'' 

\section*{Acknowledgements}

We thank Alexander N. Korotkov and Andrew N. Jordan for motivating discussions regarding the Lorentz-like structure of qubit measurement and the delayed choice nature of phase-sensitive homodyne readout of superconducting qubits. This work was partially supported by the U. S. Army Research Office under grant W911NF-22-1-0258.

\printbibliography

\appendix
\include{appendix}

\end{document}

%% file: appendix.tex
\appendix

\section*{Appendix}\label{sec:appendix-stochastics}

Here we derive the stochastic increments for both unnormalized $\hat p$ and normalized $\hat \beta$ states. Consider stochastic transformations of the form $\hat R = \exp(\lambda dt \hat F)$ where $\hat F = \hat F_{D} + \hat F_{\xi}$, and $dt \hat F_{\xi} = \sum_{j} dW_j \hat F_j$ with $(dW_j)^2 = dt$. We obtain
\begin{align}
    \hat p(t + dt) 
    = \hat R \hat p \hat R^\dagger
    = \left(1 + \lambda dt \hat F + \frac{(\lambda dt \hat F_{\xi})^2 }{2}\right) \hat p(t) \left(1 + \lambda dt \hat F^\dagger + \frac{(\lambda dt \hat F_{\xi}^\dagger)^2}{2}\right) + O(dt^{3/2}),
\end{align}
which yields the stochastic momentum increment
\begin{align}
    d \hat p 
    = \hat R \hat p \hat R^\dagger - \hat p 
    = \lambda \left(dt \left\{ \hat F_{D}, \hat p \right\} + \frac{\lambda}{2} \left\{ dt \hat F_{\xi}, \left\{ dt \hat F_{\xi}, \hat p \right\} \right\} \right) + \lambda \left\{ dt \hat F_{\xi}, \hat p \right\}+ O(dt^{3/2}) \approx d \hat p_D + d \hat p_\xi,
\end{align}
the first term deterministic, order $dt$, and the second term is stochastic, order $dW$, with anticommutator defined as $\left\{ A, B \right\} = A B + B A^\dagger$. The equation for the unnormalized qubit state $\hat s$ is identical, with $\hat H_\text{eff} = i \lambda \hat F$.

The dynamics obeyed by $\hat \beta = \hat p/\Tr \hat p$, can be derived as follows. To first order in $dW$ and $dt$, we expand
\begin{align}
     \frac{1}{\Tr \hat p + \Tr(d\hat p)} &= \frac{1}{\Tr \hat p} - \frac{\Tr(d\hat p)}{(\Tr \hat p)^2} + \frac{\Tr(d\hat p)^2}{(\Tr \hat p)^3} + O(dt^{3/2})
\end{align}
and write
\begin{align}
    \hat \beta(t + dt) 
    &= \frac{\hat p(t) + d\hat p(t)}{\Tr \hat p(t) + \Tr(d\hat p(t))}
    = \hat \beta + \frac{d\hat p}{\Tr \hat p} - \frac{\hat \beta \Tr(d\hat p)}{\Tr \hat p} - \frac{\Tr (d \hat p)}{\Tr \hat p}\left(\frac{d\hat p}{\Tr \hat p} - \frac{\hat \beta \Tr(d\hat p)}{\Tr \hat p} \right) + O(dt^{3/2}),
\end{align}
finding
\begin{align}
    d \hat \beta 
    = \frac{d\hat p}{\Tr \hat p} - \frac{\hat \beta \Tr(d\hat p)}{\Tr \hat p} - \frac{\Tr (d \hat p)}{\Tr \hat p}\bigg\vert_{dW}\left(\frac{d\hat p}{\Tr \hat p} - \frac{\hat \beta \Tr(d\hat p)}{\Tr \hat p} \right)\bigg\vert_{dW} + O(dt^{3/2}) \approx d \hat \beta_D + d \hat \beta_\xi.
\end{align}
The stochastic part of the velocity update
\begin{align}
    d \hat \beta_{\xi} = \lambda \left( \left\{dt \hat F_{\xi}, \hat \beta \right\} - \Tr\left(\left\{dt \hat F_{\xi}, \hat \beta \right\}\right) \hat \beta \right)
\end{align}
is identical to the innovation term of the stochastic master equation for the normalized density matrix $\hat \rho$. The Itô correction takes the form,
\begin{align}
    \frac{\Tr (d \hat p)}{\Tr \hat p}\bigg\vert_{dW}\left(\frac{d\hat p}{\Tr \hat p} - \frac{\hat \beta \Tr(d\hat p)}{\Tr \hat p} \right)\bigg\vert_{dW} &= \lambda^2 \Tr\left(\left\{dt \hat F_\xi, \hat \beta \right\}\right) \left( \left\{ dt \hat F_{\xi}, \hat \beta \right\} - \Tr\left(\left\{dt \hat F_\xi, \hat \beta \right\}\right) \hat \beta\right),
\end{align}
and contributes to the deterministic increment,
\begin{equation}
    d \hat \beta_{D} 
    = \lambda \left(\left\{ dt \hat F_{D}', \hat \beta \right\} - \Tr\left(\left\{ dt \hat F_{D}', \hat \beta \right\}\right) \hat \beta\right) + \lambda^2 \left(dt \hat F_{\xi} \,\hat \beta dt \hat F_{\xi}^{\dagger} - \Tr\left(dt \hat F_{\xi} \,\hat \beta dt \hat F_{\xi}^{\dagger}\right) \hat \beta\right) \\
\end{equation}
where the Itô correction modifies the effective deterministic generator of motion
\begin{align}
    dt \hat F_{D}' = dt \hat F_{D} - \lambda \Tr\left( \left\{dt \hat F_\xi, \hat \beta\right\}\right) dt \hat F_{\xi}.
\end{align}
The same dynamics govern the normalized qubit state $\hat \rho$ under identification between effective electromagnetic field and effective non-Hermitian generator, $dt \hat H_\text{eff}' = i \lambda dt \hat F' = i \lambda (dt \hat F_D' + dt \hat F_\xi)$.